\documentclass[aps,prd,preprint,showpacs,nofootinbib,floatfix]{revtex4}
\usepackage{epsfig}
\usepackage{amsmath}
\usepackage{amssymb}
\usepackage{amsfonts}
\usepackage{latexsym}
\usepackage{wasysym}
\usepackage{multirow}
\usepackage{fixmath}
\usepackage{bbm}
\usepackage{slashbox}
\newcommand{\vp}{\boldsymbol{p}}
\newcommand{\vP}{\boldsymbol{P}}
\newcommand{\vd}{\boldsymbol{d}}
\newcommand{\vk}{\boldsymbol{k}}

\newcommand{\vq}{\boldsymbol{q}}
\newcommand{\vn}{\boldsymbol{n}}
\newcommand{\vre}{\boldsymbol{r}}
\newcommand{\vv}{\boldsymbol{v}}
\newcommand{\vw}{\boldsymbol{w}}
\newcommand{\vx}{\boldsymbol{x}}
\newcommand{\vy}{\boldsymbol{y}}
\newcommand{\vz}{\boldsymbol{z}}
\newcommand{\vX}{\boldsymbol{X}}
\newcommand{\vgamma}{\boldsymbol{\gamma}}
\newcommand{\ve}{\boldsymbol{e}}

\newcommand{\vDelta}{{\boldsymbol{\Delta}}}

\def\eq{\begin{eqnarray}}
\def\en{\end{eqnarray}}

\begin{document}

\vspace*{1.0cm}

\title{Scattering phases for meson and baryon resonances on general
moving-frame lattices\\[0.5em]}

\author{M.~G\"ockeler$^{1}$, R.~Horsley$^{2}$, M. Lage$^{3}$,
  U.-G.~Mei{\ss}ner$^{3,4}$, P.E.L.~Rakow$^{5}$,\\
  A.~Rusetsky$^{3}$, G.~Schierholz$^{6}$ and J.M. Zanotti$^{7}$}

\affiliation{\vspace*{0.75cm}
\renewcommand{\baselinestretch}{1.1}\normalsize
$^1$ Institut f\"ur Theoretische Physik, Universit\"at Regensburg,
93040 Regensburg, Germany\\ \\
$^2$ School of Physics and Astronomy, University of Edinburgh, Edinburgh
EH9 3JZ, United Kingdom\\ \\
$^3$ Helmholtz-Institut f\"ur Strahlen- und Kernphysik~(Theorie)\\ and
Bethe Center for Theoretical Physics, Universit\"at
Bonn,\\ D-53115 Bonn, Germany\\ \\
$^4$ Institut f\"ur Kernphysik
(IKP-3),\\ Institute for Advanced Simulation (IAS-4), J\"ulich Center
for Hadron Physics and JARA -- High Performance Computing,\\
Forschungszentrum J\"ulich,  D-52425 J\"ulich, Germany\\ \\
$^5$ Theoretical Physics Division, Department of Mathematical Sciences,
University of Liverpool, Liverpool L69 3BX, United Kingdom\\ \\
$^6$ Deutsches Elektronen-Synchrotron DESY, D-22603 Hamburg, Germany \\ \\
$^7$ CSSM, School of Chemistry and Physics, University of Adelaide,
Adelaide SA 5005, Australia\\
\vspace*{-0.5cm}
}

\renewcommand{\baselinestretch}{1.5}\normalsize
%\author{-- QCDSF Collaboration --}

%%%%%%%%%%%%%%%%%%%%%%%%%%%%%%%%%%%%%%%%%%%%%%%%%%%%%%%%%%%%%%%%%%%%%%%

\begin{abstract}

A proposal by L\"uscher enables one to compute the scattering phases of 
elastic two-body systems from the energy levels of the lattice
Hamiltonian in a
finite volume. In this work we generalize the formalism to $S$-, $P$- and
$D$-wave meson and baryon resonances, and general total momenta. Employing  
nonvanishing momenta has several advantages, among them making a wider
range of energy levels accessible on a single lattice volume and
shifting the level crossing to smaller values of $m_\pi L$.

\end{abstract}

%\keywords{Finite Temperature QCD, Phase Diagram, Improved Wilson
%Fermions, Abelian Monopoles} 

\preprint{\vtop{\hbox{ADP-12-27/T794} \hbox{DESY 12-098}
    \hbox{Edinburgh 2012/09}\hbox{LTH 948}\hbox{June 2012}}}

\maketitle

\section{Introduction}

Most hadrons are
resonances. Lattice simulations of QCD have reached the point now   
where the masses of up and down quarks are small enough so that the
low-lying hadron resonances, such as the $\rho (770)$ and $\Delta(1232)$, can
decay via the strong interactions. 

Extracting masses and widths of unstable particles from the lattice
is made difficult by the fact that resonances cannot be identified
directly with a single energy level of the lattice
Hamiltonian. Rather, the eigenstates of the lattice Hamiltonian  
correspond to states that are characteristic of the
respective volume. In a series of papers~\cite{Luscher} L\"uscher has
derived the scattering phase shift in the infinite volume from the
volume dependence of the energy levels of the lattice Hamiltonian.

\begin{figure}[b]
\vspace*{-2.00cm}
\begin{center}
    \epsfig{file=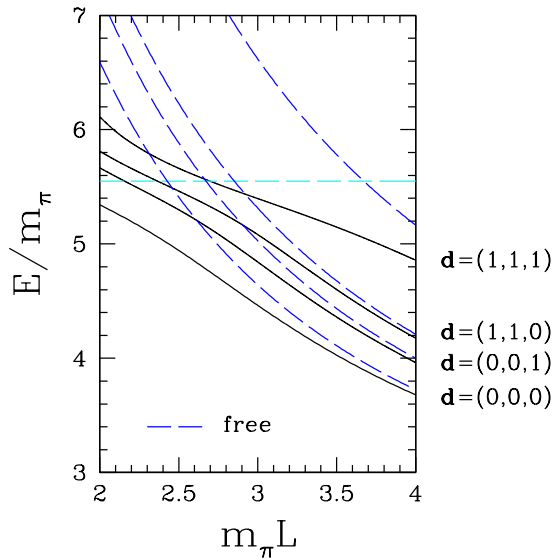,width=9.5cm}
\end{center}
\vspace*{-0.25cm}
   \caption{The ground state CM energy levels of the $\rho$ resonance
     at the physical point for various momenta 
     $\vP=(2\pi/L) \, \vd$, together with the energy levels of the
     noninteracting $\pi\pi$ system. The horizontal dashed line
     indicates the physical $m_\rho/m_\pi$ ratio.}     
\label{figr}
\end{figure}

\begin{figure}[t]
\vspace*{-1.75cm}
\begin{center}
    \epsfig{file=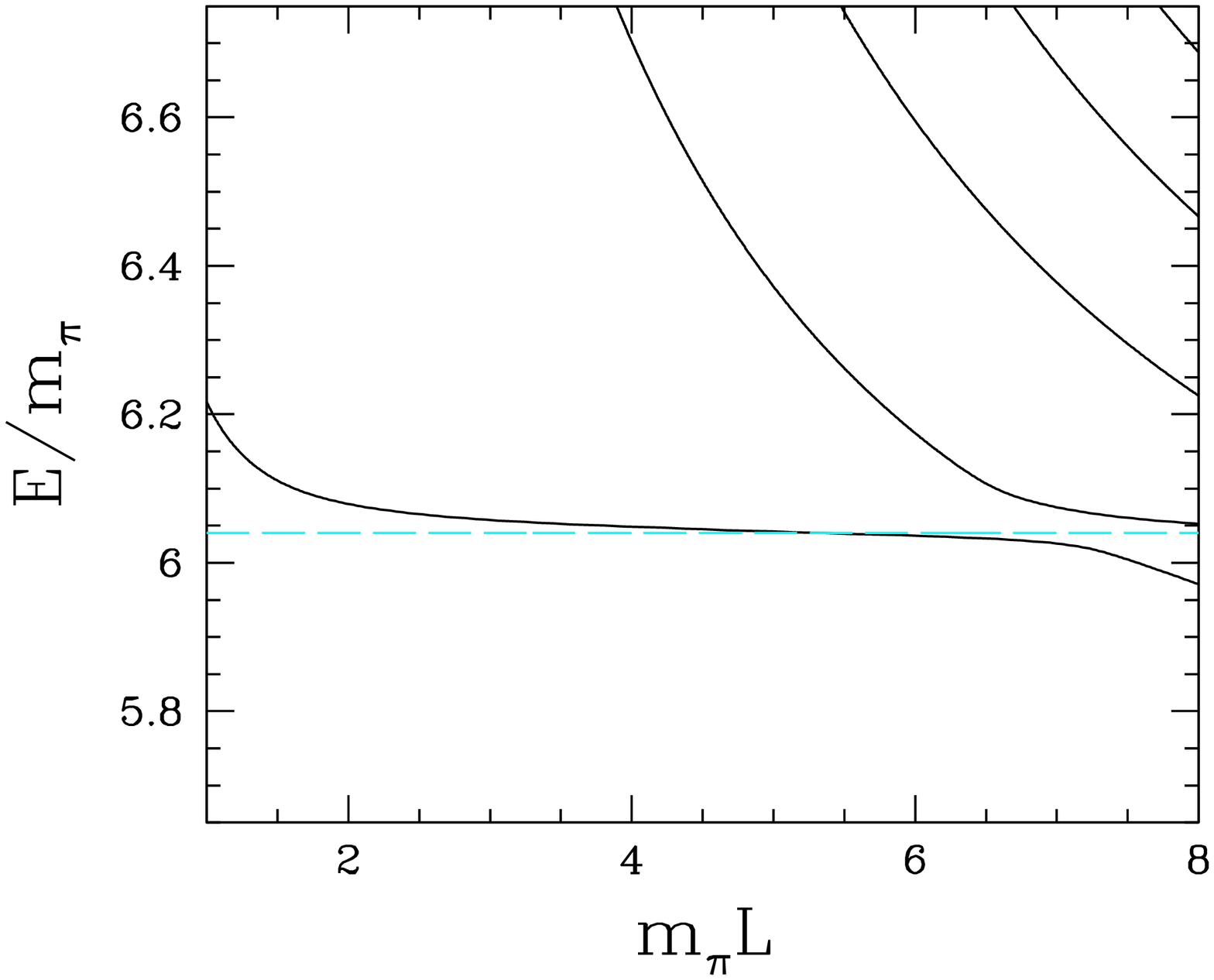,width=7.5cm}
    \epsfig{file=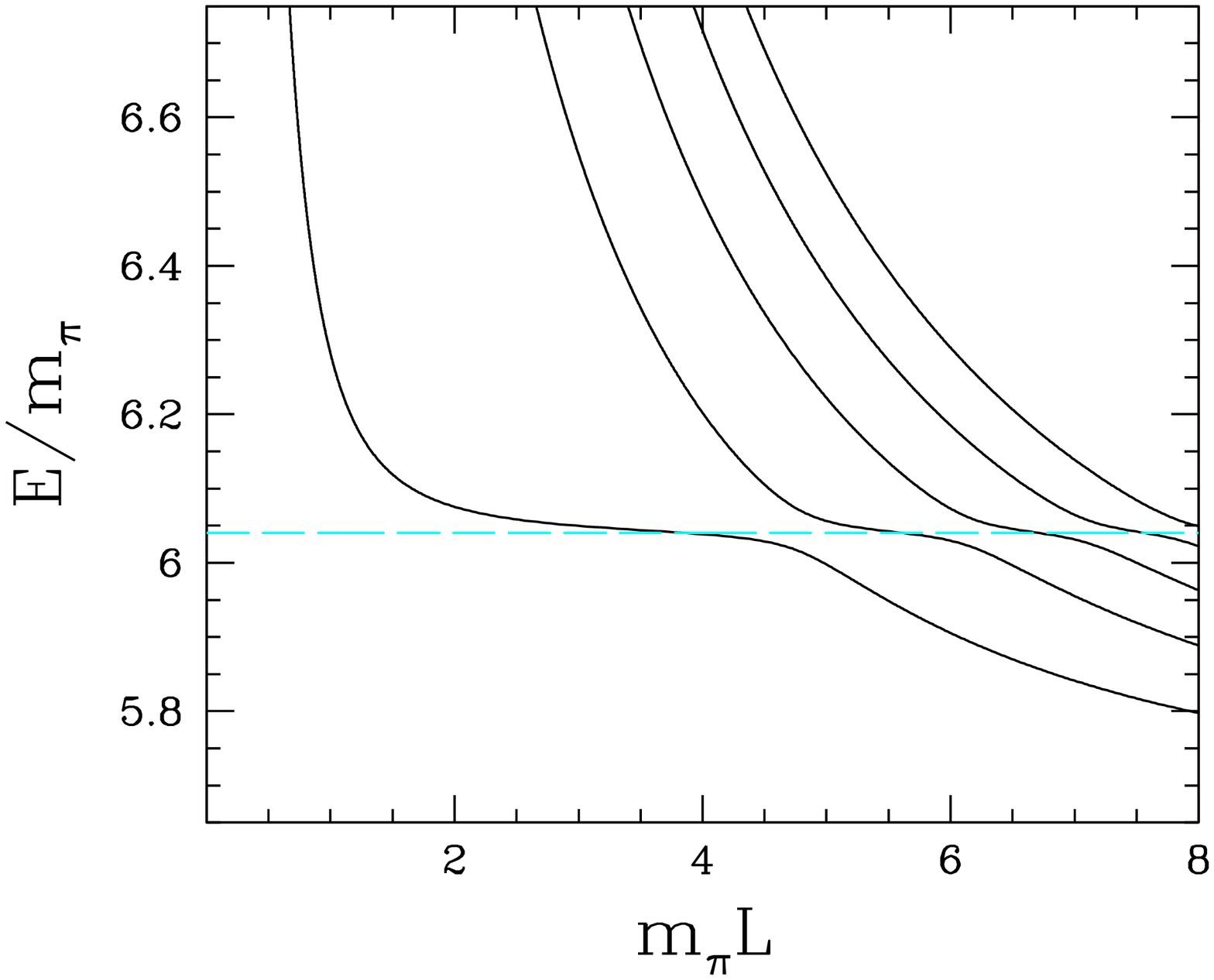,width=7.5cm}
\end{center}
\vspace*{-0.25cm}
   \caption{The expected five lowest CM energy levels of the
     $\Sigma^* (1385)$ resonance at a
     pion mass of $m_\pi = 230 \,\mbox{MeV}$~\cite{Bietenholz:2011qq}
     for zero total momentum (left) and $\vP= 2\pi/L\,\ve_3$
     (right). The dashed line indicates the anticipated
     $m_{\Sigma^*}/m_\pi$ ratio.}       
\label{figS}
\end{figure}

The original derivation was given for systems of two identical
particles with vanishing total momentum. To compute the scattering
phases for a sufficiently large set of energies on a rest-frame
lattice, one would have to repeat the calculation on several volumes,
which is computationally expensive. 

If the total momentum of the resonance is nonzero, however, a wide
variety of energy levels are becoming accessible on a single lattice
volume, as has been realized by Gottlieb and
Rummukainen~\cite{Rummukainen:1995vs}. This is illustrated in
Fig.~\ref{figr}, where we show the expected ground state energy level
of the $\rho$ resonance for several momenta  
at the physical pion mass, assuming an effective range approximation for
the scattering phase with $g_{\rho\pi\pi}=6.0$. At a lattice volume of
$m_\pi L = 2 \ldots 2.5$ the ground state energy levels of the four lowest
momenta are found to cover the resonance region already sufficiently
well. Figure~\ref{figr} tells us, furthermore, that the energy levels
of the interacting system rapidly approach the free particle energy
spectrum as $m_\pi L$ increases. For zero total momentum this limits
the region of practical use to $m_\pi L \lesssim 2.5$, while for
nonvanishing total momenta it extends to much larger values of $m_\pi
L$. 

Of particular interest to us are baryon resonances, which so far have
not been explored at all. The low-lying baryon
resonances have a much smaller phase space than, for example, the $\rho$
meson, which makes $P$-wave resonances, such as the $\Delta$ and
$\Sigma^*$, especially hard to tackle. For zero total momentum and
$O(200)\,\mbox{MeV}$ pion masses one would need volumes 
of $L \approx 6\,\mbox{fm}$ for the phase shift to cover the region
$\delta \approx \pi/2$. The reason is that the pion mass is so much 
smaller than the 
mass of the nucleon and the $\Lambda$. Not so for nonzero total
momenta though, which allows the pion to have zero momentum. In this
case the avoided level crossing of the energy levels is shifted   
towards much smaller values of $m_\pi L$. This is illustrated in 
Fig.~\ref{figS}, where we compare the expected five lowest energy
levels of the $\Sigma^*$ resonance, that decays to $\Lambda \pi$, 
for zero and nonzero total momentum, assuming again an effective range
approximation for the scattering phase with $g_{\Sigma^* \Lambda\pi} =
9.2$.   

Considering the difficulty of computing the properties of resonances on
the lattice, how was ist possible that the mass spectrum of the 
pseudoscalar and vector meson octet and the baryon decuplet
computed in~\cite{Durr:2008zz,Bietenholz:2011qq}, using standard
techniques, agreed so well with experiment? The answer is given in
Fig.~\ref{figS}. In smaller, favorable volumes the ground state energy
may agree well with the resonance mass over a wide range of $m_\pi
L$. In larger volumes the ground state energy will approach the 
energy level of two free particles though. 

Gottlieb and Rummukainen have extended L\"uscher's work on meson
resonances to nonvanishing total momentum $\vP=(2\pi/L)\,\ve_3$. Their
work was generalized further to two-body systems of arbitrary mass by
Davoudi and Savage~\cite{Davoudi:2011md} and
Fu~\cite{Fu:2011xz}. Recently, Feng {\it et al.}~\cite{Feng:2011ah}
have derived finite size formulae for the next higher momentum
$\vP=(2\pi/L)\,(\ve_1 + \ve_2)$, which has been generalized again to
particles of arbitrary masses by Leskovec and
Prelovsek~\cite{Leskovec:2012gb}. In the case of unequal masses and
nonvanishing momenta the extraction of phase shifts from the energy
levels of the lattice Hamiltonian proves difficult, because the partial
waves of the individual scattering channels will mix in
general. Strategies of how to overcome this problem have been
discussed by D\"oring {\it et al.}~\cite{Doring:2012eu} in the
framework of unitarized chiral perturbation theory,
which is equivalent to L\"uscher's approach in the
large-$L$ limit. In  
this work we shall derive phase shift formulae for meson
and baryon resonances for total momenta proportional to 
$\vP=(2\pi/L)\,\ve_3$, $\vP=(2\pi/L)\,(\ve_1 + \ve_2)$ and
$\vP=(2\pi/L)\,(\ve_1 + \ve_2 + \ve_3)$, including rotations of
$\vP$. Our formulae will cover all two-body $S$-, $P$- 
and $D$-wave meson and baryon resonances.  

Knowing the scattering phase shifts for general total momenta, among
others, we
will be able to extract a great variety of other hadronic observables,
including elastic and transition form factors of unstable particles,
such as the $\rho$ form factor and the $\Delta$ to nucleon
electromagnetic transition form factors.  

The paper is organized as follows. Section~\ref{sec2} deals with the
kinematics of two-particle states on the periodic lattice. In
Sec.~\ref{sec3} we discuss the solutions of the Helmholtz equation for
noninteracting and interacting particles. The Lorentz boost from the
laboratory frame to the center of mass frame deforms the cubic lattice,
and only some subgroups (little groups) of the original cubic point
symmetry group remain. In Sec.~\ref{sec4} we discuss the symmetry properties
of the various center of mass frames, including the representations of
the little groups. This is followed by
the reduction of the phase shift formulae according to spin, angular
momentum and representation in Sec.~\ref{sec5}. In Sec.~\ref{secex}
we give explicit expressions for the phase shifts of the $\rho$,
$\Delta$ and $N^\star(1440)$ (Roper) resonances, and in Sec.~\ref{sec6}
we give a sample of operators that transform according to some of the
prominent representations. Finally, in Sec.~\ref{sec7} we conclude.

\clearpage
\section{Two-particle kinematics on a moving-frame lattice}
\label{sec2}

In this section we discuss the kinematical properties of two
noninteracting particles of mass $m_1$ and $m_2$ in a cubic box of
length $L$ with periodic boundary conditions. Twisted boundary
conditions will be discussed elsewhere.  

Let us first consider the lattice or laboratory (L) frame. We denote
the 3-momenta of the individual particles by $\vp_1, \vp_2$. The total
momentum is denoted by $\vP =\vp_1 +  \vp_2$. The energy of 
two free particles is given by 
\begin{equation}
W=\sqrt{\vp_1^2+m_1^2}+\sqrt{\vp_2^2+m_2^2} \,. 
\label{wp}
\end{equation}
The lattice momenta $\vp_i$ are quantized to
\begin{equation}
\vp_i = \frac{2\pi}{L} \vn_i \,, \quad \vn_i \in
  \mathbb{Z}^3 \,,
\end{equation} 
and, similarly,
\begin{equation}
\vP = \frac{2\pi}{L} \vd \,, \quad \vd \in \mathbb{Z}^3 \,.
\end{equation} 

Next, we consider the center-of-mass (CM) frame, which is moving
with velocity
\begin{equation}
\vv = \frac{\vP}{W}\,, \quad v = |\vv| \,
\end{equation}
in the laboratory frame.
We denote the CM (relative) momentum by $\vk$ and the energy by
$E$. Momentum and energy are obtained by a standard Lorentz
transformation,  
\begin{equation}
%\vk = \underbrace{\vgamma\left( \vp_1 - \vv
%  \sqrt{\vp_1^2+m_1^2}\right)}_{\displaystyle \vp_1^{\rm CM}} = -
%\,\underbrace{\vgamma \left( \vp_2 - \vv 
\vk = \vgamma\left( \vp_1 - \vv
  \sqrt{\vp_1^2+m_1^2}\right) = -
\,\vgamma \left( \vp_2 - \vv 
\sqrt{\vp_2^2+m_2^2}\right) \,,
\label{lor}
\end{equation}
where 
\begin{equation}
\vgamma \vp = \gamma \vp_\parallel + \vp_\perp \,, \quad \gamma =
\frac{1}{\sqrt{1-v^2}} = \frac{W}{E}
\end{equation}
and 
\begin{equation}
\vp_\parallel = \vv \;\frac{\vv \vp}{v^2} \,, \quad  \vp_\perp = \vp
- \vp_\parallel \,.
\end{equation}
Laboratory and CM frame energies are related by
\begin{equation}
W=\sqrt{\vP^2+E^2} \,, 
\label{wpe}
\end{equation}
where
\begin{equation}
E= E_1 + E_2 =\sqrt{\vk^2+m_1^2}+\sqrt{\vk^2+m_2^2} \,, \quad 
%\end{equation}
%and
%\begin{equation}
\vk^2= \frac{\left(E^2-(m_1^2+m_2^2)\right)^2 - 4m_1^2 m_2^2}{4 E^2}
\,.
\label{ke}
\end{equation}
Defining
\begin{equation}
\vp = \frac{1}{2}\, \left(\vp_1-\vp_2\right) = \vp_1 - \frac{1}{2}\,
\vP = -\vp_2 + \frac{1}{2}\,\vP \,,
\end{equation}
and expressing the laboratory frame energies in (\ref{lor}) by their
CM counterparts, the CM momentum $\vk$ can be rewritten as
\begin{equation}
%\begin{split}
%\vk &= \gamma\phantom{^{-1}} \, \frac{m_2^2-m_1^2}{W^2}\,\frac{1}{2}
%\vP+ \vgamma^{-1}\vre \\[0.25em]
\vk = 
\vgamma^{-1} \vp - \gamma^{-1}
\,\frac{m_1^2-m_2^2}{E^2}\,\frac{1}{2}\, \vP
\,,\quad \vgamma^{-1}\vp =\gamma^{-1} \vp_\parallel + \vp_\perp \,.
%\end{split}
\end{equation}
This results in the quantization condition
\begin{equation}
\vk \in \Gamma_\vDelta
\end{equation}
with
\begin{equation}
\Gamma_\vDelta= \left\{\vk\, \Big|\, \vk=\frac{2\pi}{L} \vgamma^{-1}
\left( \vn - \frac{1}{2}\, \vDelta\right) \,, \, \vn \in
\mathbb{Z}^3\right\}\,, \quad  \vDelta =
\vd\,\left(1+\frac{m_1^2-m_2^2}{E^2}\right)  \,.
\label{qcon}
\end{equation}
  
%In the interacting case the energy spectrum is still given by the
%dispersion relations (\ref{wpe}) and (\ref{ke}), but with $\vk^2$ no
%longer being quantized. 

\section{Solutions of the Helmholtz equation}
\label{sec3}

To compute the scattering phases of the interacting two-particle system,
we need to discuss the solutions of the Helmholtz
equation~\cite{Luscher2} in the CM frame first. 
 
In the laboratory frame the two-particle state is described by the wave 
function $\psi_L(x_1;x_2)$, where $x_1=(x^0_1,\vx_1)$,
$x_2=(x^0_2,\vx_2)$ are the space-time coordinates in Minkowski
space. For the moment we restrict ourselves to particles of spin zero.
The wave function can then be written 
\begin{equation}
\psi_L(x_1;x_2) = e^{-i(Wt - \vP\vX)}\, \phi_L(x_0,\vx)
\label{wf}
\end{equation}
with
\begin{equation}
\begin{split}
\vX &= \frac{m_1\vx_1 + m_2\vx_2}{m_1 + m_2}\,,\quad
\vx = \vx_1 - \vx_2 \,,\\
t &= \frac{m_1 x_1^0 + m_2 x_2^0}{m_1 + m_2}\,,\quad x^0=x_1^0-x_2^0\,.
\end{split}
\label{tt}
\end{equation}
We are interested in the case where both particles have equal
time coordinates, $x_1^0 = x_2^0 = t$.

We denote the space-time separation in the CM frame by
$r=(r^0,\vre)$. The transformation from the laboratory frame to the CM
is given by 
\begin{equation}
\left(\begin{tabular}{c}
$r^0$\\$\vre$
\end{tabular}\right) =
\left(\begin{tabular}{cc}
$\gamma$ \,& \,$\gamma\,\vv$\\
$\vgamma \vv$ \,& \,$\vgamma$
\end{tabular}\right)\,
\left(\begin{tabular}{c}
$0$\\$\vx$
\end{tabular}\right)  
\end{equation}
with 
\begin{equation}
\phi_L(0,\vx) = \phi_{CM}(r^0,\vre) \,.
\label{wfcm}
\end{equation}
In the case of unequal masses, $m_1 \neq m_2$, the (relative) time
coordinate $r^0$ is no longer zero, even though $x^0$ is. 

\subsection{Noninteracting particles}

For noninteracting particles the CM frame wave function obeys the
equation of motion 
\begin{equation}
\left( -\nabla^2_{r^0} + \pmb{\nabla}^2_{\vre} + \left(E^2-
(m_1+m_2)^2\right) \frac{m_1 m_2}{(m_1 +m_2)^2}\right)\,
\phi_{CM}(r^0,\vre) = 0
\label{mh}
\end{equation}
with
\begin{equation}
\left( -i\,\nabla_{r^0} - \frac{E_1 m_2 - E_2 m_1}{m_1 +m_2}\right)\,
\phi_{CM}(r^0,\vre) = 0 \,.
\label{mmh}
\end{equation}
Equations (\ref{mh}) and (\ref{mmh}) follow directly from the
Klein-Gordon equations of the individual particles. Writing 
\begin{equation}
\phi_{CM}(r^0,\vre) = e^{i\, \frac{E_1 m_2 - E_2 m_1}{m_1 +m_2} \,
  r^0}\, \phi_{CM}(0,\vre) \,,%= e^{i\, \left(\frac{m_1^2 - m_2^2}{2E} -
  %\frac{m_1 -m_2}{m_1 +m_2}\right) \,
  %r^0}\, \phi_{CM}(0,\vre)
\label{wfr}
\end{equation} 
the time dependence can be factored out, and we obtain the Helmholtz
equation 
\begin{equation}
\left( \pmb{\nabla}^2_{\vre} + \vk^2 \right) \, \phi_{CM}(\vre) = 0
\label{he}
\end{equation}
with $\phi_{CM}(\vre) = \phi_{CM}(0,\vre)$, and $\vk$ given by
(\ref{qcon}). 
%symmetric. In the case of noninteracting 
%particles $\vk$ is given by (\ref{qcon}), while in the interacting
%case $\vk^2$ is the solution of a nonlinear equation involving the
%scattering phase shift. 

The laboratory frame wave function is periodic under spatial
translations  
\begin{equation}
\psi_L(x_1^0,\vx_1;x_2^0,\vx_2) =
\psi_L(x_1^0,\vx_1+\vn_1L;x_2^0,\vx_2+\vn_2L)\,, \quad \vn_{1,2} \in
\mathbb{Z}^3. 
\end{equation}
Equations (\ref{wf}), (\ref{wfcm}) and (\ref{wfr}) together give
\begin{equation}
\psi_L(0,\vx_1;0,\vx_2) = e^{i \,\left( \vP\vX + \frac{E_1m_2-E_2m_1}{E
    (m_1+m_2)} \vP \vx \right)}\,
  \phi_{CM}(\vre) \,,
\end{equation}
where we have inserted $r^0=\gamma\, \vv \vx = \vP \vx/E$. This leads to the
periodicity relation for the CM wave function
\begin{equation}
\phi_{CM}(\vre) = e^{-i\, \pi \,\vn\vDelta}\, \phi_{CM}(\vre+\vgamma\vn
L) 
\label{bc}
\end{equation}
with $\vn=\vn_1-\vn_2$. 
%Note that (\ref{bc}) is formally invariant
%under the interchange of particles $1 \leftrightarrow 2$. 
While for
equal masses $\phi_{CM}(\vre)$ is either periodic ($\vn$ even) or
antiperiodic ($\vn$ odd) with period $\gamma L$, this is no longer so
for $m_1 \neq m_2$. In this case the CM wave function picks up a
complex phase factor $e^{- i\,\pi\,\vn\vDelta}$ when crossing the spatial
boundary. We call this attribute $\vDelta$-periodic. 

\subsection{Interacting particles}

Let us now turn to the interacting case. We assume that the two-body
interaction has finite range and vanishes outside the region $|\vre| >
R$ with $L \gg 2 R$. In the exterior region $\phi_{CM}(\vre)$ satisfies
the Helmholtz equation   
\begin{equation}
\left( \pmb{\nabla}^2_{\vre} + \vk^2 \right) \, \phi_{CM}(\vre) = 0 
\label{he2}
\end{equation}
with
\begin{equation}
\vk^2= \frac{\left(E^2-(m_1^2+m_2^2)\right)^2 - 4m_1^2 m_2^2}{4
  E^2} \equiv k^2 \,, 
\end{equation} 
where $E$ now are the energy levels of the
interacting system. 

We are now looking for solutions of the Helmholtz
equation (\ref{he2}). The (singular) case $\vk \in \Gamma_\vDelta$
requires a separate discussion, which we shall omit here. The Green
function \begin{equation}
G^\vDelta(\vre,k^2)= \gamma^{-1} L^{-3} \sum_{\vp \in \Gamma_\vDelta}
\frac{e^{\,i\, \vp \vre}}{\vp^2 - k^2} 
\label{gf}
\end{equation}
is such a solution. An appropriate basis of  
solutions of the Helmholtz equation is obtained from (\ref{gf}) by
\begin{equation}
G^\vDelta_{l m}(\vre,k^2)= \mathcal{Y}_{l m}(\pmb{\nabla})\,
G^\vDelta(\vre,k^2) \,,
\end{equation}
where 
\begin{equation}
\mathcal{Y}_{l m}(\vre) = |\vre|^l \, Y_{l m}(\hat{\vre}) \,, \quad
\hat{\vre} = \frac{\vre}{|\vre|} \,.
\end{equation}
Obviously, $G^\vDelta(\vre,k^2)$ and $G^\vDelta_{l m}(\vre,k^2)$ are
$\vDelta$-periodic. The CM wave function can then be expanded as
\begin{equation}
\phi_{CM}(\vre) = \sum_{l,m} c_{l m}\, G^\vDelta_{l m}(\vre,k^2) \,,
\label{ggf}
\end{equation}
which may be interpreted as a partial wave expansion. The
functions $G^\vDelta_{l m}$ can be expanded in spherical harmonics
$Y_{lm}(\theta,\varphi)$ and spherical Bessel functions~\cite{Messiah}
$n_l(kr)$, $j_l(kr)$ 
\begin{equation}
G^\vDelta_{l m}(\vre,k^2) = \frac{(-1)^l\, k^{l+1}}{4\pi}
\left[n_l(kr)\, Y_{lm}(\theta,\varphi) + \sum_{l^\prime =0}^\infty
  \sum_{m^\prime = -l^\prime}^{l^\prime} M_{l m, l^\prime m^\prime}^\vDelta\,
  j_{l^\prime}(kr)\, Y_{l^\prime m^\prime}(\theta,\varphi)\right]
\end{equation}
with
\begin{equation}
M_{l m, l^\prime
  m^\prime}^\vDelta=\frac{(-1)^l\,\gamma^{-1}}{\pi^{3/2}}
\sum_{j=|l-l^\prime|}^{l+l^\prime}\, \sum_{s=-j}^j \frac{i^j}{q^{j+1}}\,
  Z_{j s}^\vDelta (1,q^2)^*\, C_{lm,js,l^\prime m^\prime} \,, \quad
q=\frac{k L}{2\pi} \,,
\label{mz}
\end{equation}
where $r=|\vre|, \theta$ and $\varphi$ are the polar coordinates of
$\vre$. The generalized zeta function $Z_{j s}^\vDelta(1,q^2)$ is
obtained from 
\begin{equation}
Z_{j s}^\vDelta (\delta,q^2) = \sum_{\vz \in P_\vDelta}
\frac{\mathcal{Y}_{js}(\vz)}{(\vz^2 - q^2)^\delta} 
\label{yy}
\end{equation}
with
\begin{equation}
P_\vDelta = \left\{\vz\, \Big|\, \vz= \vgamma^{-1} \left( \vn
- \frac{1}{2}\, \vDelta\right) \,, \, \vn \in
\mathbb{Z}^3\right\} 
\label{gg}
\end{equation}
by analytic continuation $\delta \rightarrow 1$. The coefficient
$C_{lm,js,l^\prime 
  m^\prime}$ can be expressed in terms of Wigner $3j$-symbols
\begin{equation}
C_{lm,js,l^\prime  m^\prime} = (-1)^{m^\prime}\, i^{l-j+l^\prime}
\sqrt{(2l+1)(2j+1)(2l^\prime +1)} \left(\begin{tabular}{ccr} $l$ & $j$
  & $l^\prime$ \\ $m$ & $s$ & $-m^\prime$ \end{tabular}\right) \,
\left(\begin{tabular}{ccc} $l$ & $j$
  & $l^\prime$ \\ $0$ & $0$ & $0$ \end{tabular}\right) \,.
\end{equation}
In Appendix A we give $Z_{j s}^\vDelta(1,q^2)$ for
arbitrary values of $j$ and $s$. 

It is easily seen that
\begin{equation}
\begin{split}
&\left\{\vz\, \Big|\, \vz= \vgamma^{-1} \left( \vn
- \frac{1}{2}\, \vDelta(m_1,m_2) \right) \,, \, \vn \in
\mathbb{Z}^3\right\}\\[0.5em]
&= \left\{-\vz\, \Big|\, \vz= \vgamma^{-1} \left( \vn
- \frac{1}{2}\, \vDelta(m_2,m_1) \right) \,, \, \vn \in
\mathbb{Z}^3\right\} \,.
\end{split}
\end{equation}
This results in
\begin{equation}
Z_{j s}^{\vDelta(m_1,m_2)}(\delta,q^2) = (-1)^j\, Z_{j
  s}^{\vDelta(m_2,m_1)}(\delta,q^2) 
\label{zzz}
\end{equation}
and
\begin{equation}
M_{lm,l^\prime m^\prime}^{\vDelta(m_1,m_2)}=(-1)^{l+l^\prime} \,
M_{lm,l^\prime m^\prime}^{\vDelta(m_2,m_1)} \,,
\end{equation}
as $l+j+l^\prime = \mbox{even}$. Both $M_{lm,l^\prime
  m^\prime}^{\vDelta(m_1,m_2)}$ and $M_{lm,l^\prime 
  m^\prime}^{\vDelta(m_2,m_1)}$ have the same determinant and lead to
the same results, so that the 
order of $m_1$ and $m_2$ does not matter. 

In the literature one often finds the
expression~\cite{Rummukainen:1995vs,Leskovec:2012gb} 
\begin{equation}
M_{l m, l^\prime
  m^\prime}^\vDelta=\frac{(-1)^l\,\gamma^{-1}}{\pi^{3/2}}
\sum_{j=|l-l^\prime|}^{l+l^\prime}\, \sum_{s=-j}^j \frac{i^j}{q^{j+1}}\,
  Z_{j s}^\vDelta (1,q^2)\, C_{lm,js,l^\prime m^\prime} \,.
\label{mzz}
\end{equation}
Though not quite correct in general, it leads to the same results for the phase
shifts as the matrix
(\ref{mz}). Indeed, if we denote (\ref{mzz}) by $\tilde{M}$, we find 
$\tilde{M}_{lm,l^\prime m^\prime}=(-1)^{l+l^\prime}\,M_{lm,l^\prime
  m^\prime}^*$, which has the same determinant as $M$ (see the
equations for the phase shifts given in (\ref{boson}) and
(\ref{baryon}) below). 
In the following we shall use the short-hand notation
\begin{equation}
w_{l m} = \frac{1}{\pi^{3/2} \sqrt{2l+1}} \gamma^{-1} q^{-l-1}\,
  Z_{lm}^\vDelta (1,q^2) \,.
\label{wz}
\end{equation}

So far we have considered spinless particles only. Let us now assume
that one of the particles carries spin $S$. In the outer region
$|\vre| > R$, which we are concerned with here, the spin operator
$\hat{S}$ commutes with the Hamiltonian. The spin-dependent part of
the wave function can thus be factored out. As we are mainly interested
in meson-baryon resonances, we consider $S=1/2$. In this case we have
\begin{equation}
\phi_{CM}(\vre) = \sum_{\substack{J,\mu \\ l,m,\sigma}} \langle
lm,\frac{1}{2}\sigma|J\mu\rangle\, c_{lm}\,
G_{lm}^\vDelta(\vre,q^2)\,\chi_\sigma^{\frac{1}{2}} \,,
\end{equation}
where $\chi^{\frac{1}{2}}_\sigma$ is the two-component baryon
spinor. This amounts to an expansion of the CM wave function in terms
of spin sperical harmonics
\begin{equation}
Y_{J l \mu}= \sum_{m,\sigma} \langle lm,\frac{1}{2}\sigma|J\mu\rangle \,
Y_{lm}\,\chi_\sigma^{\frac{1}{2}} \,. 
\end{equation}
In this basis the matrix $M^\vDelta$ reads
\begin{equation}
M_{Jl\mu,J^\prime l^\prime \mu^\prime}^\vDelta =
\sum_{\substack{m,\sigma\\m^\prime,\sigma^\prime}} \langle
lm,\frac{1}{2}\sigma|J\mu\rangle\,\langle
l^\prime m^\prime,\frac{1}{2}\sigma^\prime|J^\prime\mu^\prime \rangle
\, M_{l m, l^\prime m^\prime}^\vDelta\,.
\end{equation}

\section{Symmetry properties}
\label{sec4}

The Lorentz boost deforms the cubic box to a parellelepiped, in
which the length scale parallel to the direction of the boost vector
is multiplied by $\gamma$, whereas the perpendicular length scale is
left unchanged.

\subsection{Boost vectors}

\begin{figure}[t]
\vspace*{-0.25cm}
\begin{center}
%\epsfig{file=par1.ps,width=7cm,clip=}
%\raisebox{0.07cm}{\epsfig{file=par2.ps,width=7.97cm,clip=}}
\epsfig{file=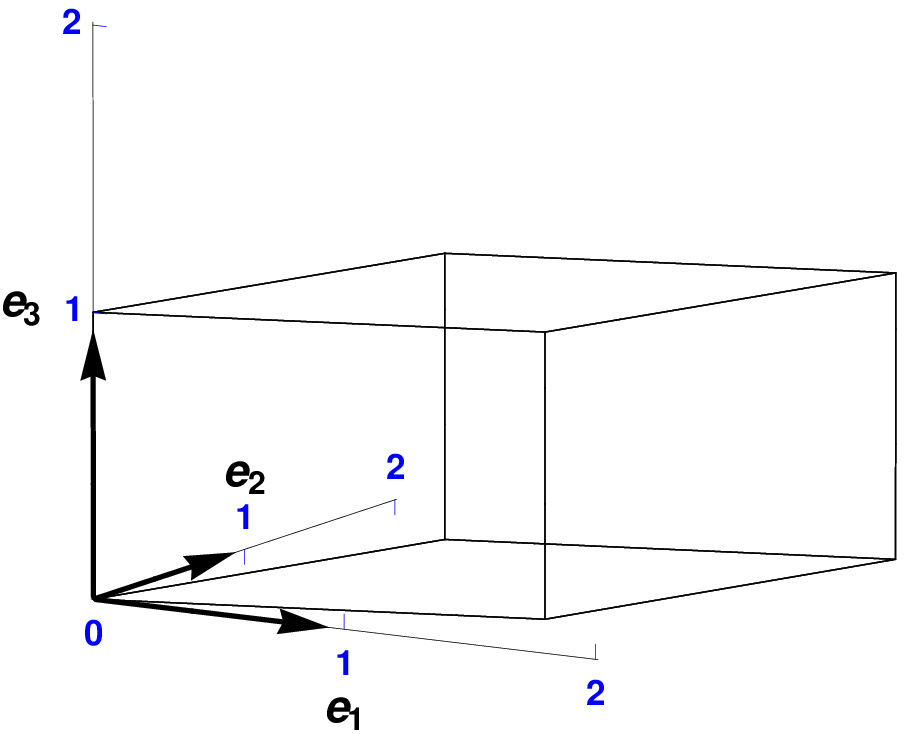,width=7cm,clip=}\hspace*{1.0cm}
\epsfig{file=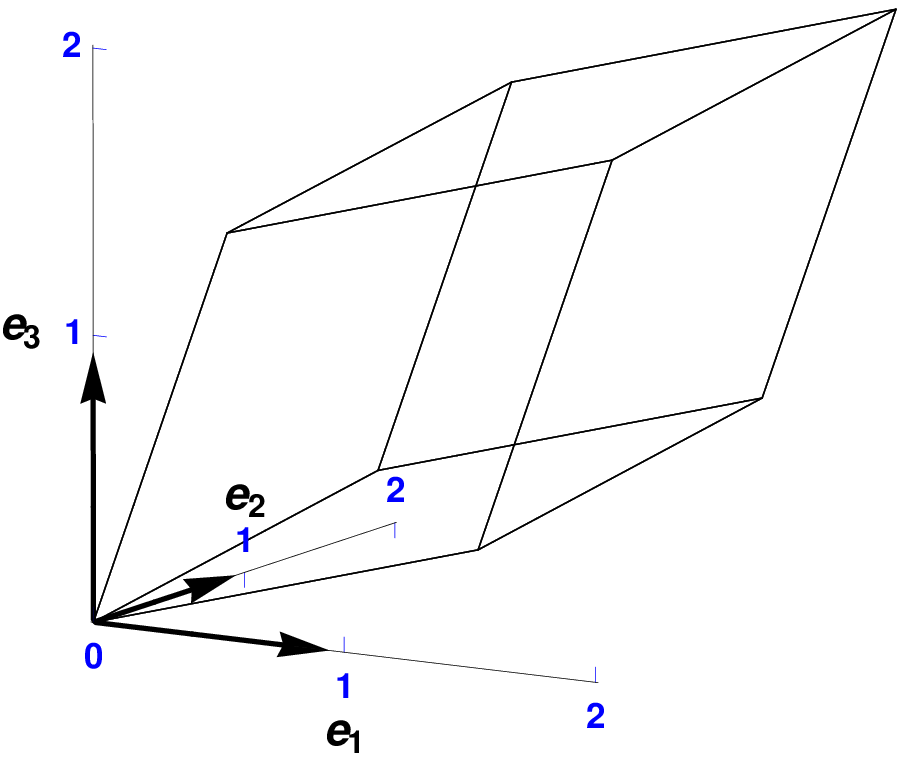,width=7cm,clip=}
\end{center}
%\vspace*{-0.75cm}
\caption{A cubic box of unit length deformed to a parallelepiped with
  $\vd=(\ve_1+\ve_2)$ (left panel) and $\vd=(\ve_1+\ve_2+\ve_3)$
  (right panel) for $\gamma=2$.} 
\label{para}
\end{figure}

We will consider boost vectors
\begin{equation}
\vd=(d_1,d_2,d_3)\,,\quad d_i=0, \pm 1
\label{v}
\end{equation}
and integer multiples $n\,\vd$, $n\in \mathbb{Z}$ thereof. For
that purpose it is sufficient to 
consider 
\begin{equation}
\begin{tabular}{lcl}
$\vd = (0,0,0)\, \equiv\, {\bf 0}$\,\, &\,,&\,\,
$\vd = (0,0,1)\, \equiv\, \ve_3$\,\,,\\[0.5em]
$\vd = (1,1,0)\, \equiv\, \ve_1+\ve_2$\,\, & \,,&\,\,
$\vd = (1,1,1)\, \equiv\, \ve_1+\ve_2+\ve_3$\,\,.
\end{tabular}
\label{d}
\end{equation}
The boost vectors (\ref{v}) can be transformed into one of the boost
vectors (\ref{d}) by a global rotation, which will leave our final
results unchanged. Results for multiples of (\ref{d}) are obtained by
simply replacing $\vd$ by $n\,\vd$ in the formulae to follow. 
In Fig.~\ref{para} we show two examples of the deformation of the
cubic box. 

\subsection{Properties of the functions $\mathbold{w_{lm}}$}

In the following we shall use the shorthand notation
\eq
\vw_{l}=(w_{l\,l},w_{l\,l-1},\cdots w_{l\,-l+1},w_{l\,-l})\, .
\en
As a result of (\ref{zzz}),
%$Z^{-\bf d}_{js}(1;q^2)\neq Z^{\bf d}_{js}(1;q^2)$
$w_{lm}$ is no longer zero for odd values of $l$ in the case of
unequal masses. In general we have
\eq
Z^{\vDelta}_{l-m}(1;q^2)=(-1)^m  Z^{\vDelta}_{lm}(1;q^2)^* \,.
\en

\subsubsection{The case $\vd=(0,0,1)$}

%\begin{itemize}
%\item
In this case the system is symmetric under rotations
around $\ve_3$ by $\pi/2$, which leads to
\eq
Z^{\vDelta}_{lm}(1;q^2)=0\,,\quad \mbox{for}\; m\neq 0 \mod 4 \,.
\en
Furthermore, the system is symmetric under the interchange of axes $1
\leftrightarrow 2$, as well as reflections $\ve_{1,2} \rightarrow
-\ve_{1,2}$. This leaves us with the following elements
\begin{equation*}
\vspace*{0.35cm}
\begin{tabular}{|c| c|}
\hline
$l$ & $\vw_l$ \\
\hline
0 & $(w_{00})$ \\
1 & $(0,w_{10},0)$\\
2 & $(0,0,w_{20},0,0)$\\
3 & $(0,0,0,w_{30},0,0,0)$\\
4 & $(w_{44},0,0,0,w_{40},0,0,0,w_{44})$\\
\hline
\end{tabular}
\end{equation*}

\subsubsection{The case $\vd=(1,1,0)$}

In this case the system is symmetric under the interchange of 
the axes $1
\leftrightarrow 2$. Furthermore, the system is symmetric under
reflections $\ve_3 \rightarrow -\ve_3$. This leaves us with the
following elements 
\begin{equation*}
\vspace*{0.35cm}
\begin{tabular}{|c| c|}
\hline
$l$ & $\vw_l$ \\
\hline
0 & $(w_{00})$\\
1 & $\sqrt{2}\,{\rm Re}\,w_{11}\,\left(e^{i\pi/4},0,-e^{-i\pi/4}\right)$\\
2 & $(w_{22},0,w_{20},0,-w_{22})$\\
3 & $\sqrt{2}\,\left(e^{-i\pi/4}{\rm Re}\,w_{33},0,e^{i\pi/4}{\rm Re}\,w_{31},0,-e^{-i\pi/4}{\rm Re}\,w_{31},0,-e^{i\pi/4}{\rm Re}\,w_{33}\right)$\\
4 & $(w_{44},0,w_{42},0,w_{40},0,-w_{42},0,w_{44})$\\
\hline
\end{tabular}
\end{equation*}

\subsubsection{The case $\vd=(1,1,1)$}

In this case the only symmetry that is left is the symmetry under
cyclic permutation $1 \rightarrow 2 \rightarrow 3 \rightarrow 1$. This
leaves us with the following elements
\begin{equation*}
\vspace*{0.35cm}
\begin{tabular}{|c|c|}
\hline
$l$ & $\vw_l$ \\
\hline
0 & $(w_{00})$\\
1 & $w_{10}\,\left(-e^{i\pi/4},1,e^{-i\pi/4}\right)$\\
2 & $w_{22}\,\left(1,-\sqrt{2}e^{-i\pi/4},0,-\sqrt{2}e^{i\pi/4},-1\right)$\\
3 & $\left(-\frac{\sqrt{10}}{4}\,e^{-i\pi/4}w_{30},w_{32},\frac{\sqrt{6}}{4}\,e^{i\pi/4}\,w_{30},w_{30},-\frac{\sqrt{6}}{4}\,e^{-i\pi/4}\,w_{30},
-w_{32},\frac{\sqrt{10}}{4}\,e^{i\pi/4}w_{30}\right)$\\
4 & $\left(\frac{\sqrt{70}}{14}\,w_{40},
-\frac{\sqrt{7}}{2}\,e^{i\pi/4}w_{42},w_{42},
\frac{1}{2}\,e^{-i\pi/4}w_{42},w_{40},
\frac{1}{2}\,e^{i\pi/4}w_{42},-w_{42},
-\frac{\sqrt{7}}{2}\,e^{-i\pi/4}w_{42},\frac{\sqrt{70}}{14}\,w_{40}\right)$\\
\hline
\end{tabular}
\end{equation*}

\subsection{Irreducible representations of the little groups}

In the CM frame, the symmetry group of the cubic lattice is 
the cubic group $O$ for particles with  
integer spin, and its double cover group $^2O$
for  particles with half-integer spin. The group $O$
consists of 24 elements $R_i$, {\it i.e.} rotation matrices, which are
characterized by the axis $\vn^{(i)}$ and angle $\omega_i$
of rotation (with $i=1, \cdots, 24$). The rotation matrices are given
by    
\eq
(R_i)_{\alpha\beta}=\cos \omega_i\,\delta_{\alpha\beta}
+(1-\cos\omega_i)\,n^{(i)}_\alpha n^{(i)}_\beta
-\sin\omega_i\,\epsilon_{\alpha\beta\gamma}\,n^{(i)}_\gamma\,, \quad
\alpha,\beta,\gamma=1,2,3 \, .
\en
The 24 elements of $O$ fall into five different conjugacy
classes. They are listed in Table~\ref{Table:RotationsOfO}.
The group $^2O$ has 48 elements $R_i$. As for $O$, they are characterized
by the axis $\vn^{(i)}$ and angle $\omega_i$ (with $i=1, \cdots, 48$
now). The 48 elements of $^2O$ fall into eight different conjugacy
classes. They are listed in Table~\ref{Table:RotationsOf2O}. 

\begin{table}[t]
\begin{center}
\begin{tabular}{|c|c|c|c|}\hline
 Class &$i$   &$\vn^{(i)}$  & $\omega_i$    \\
\hline
 $I$   & 1    & any       & $0$             \\ \hline
 \multirow{8}{*}{$8C_3$}& 2    &$(1,1,1)/\sqrt{3}$  & $-2\pi/3$       \\
       & 3    &$(1,1,1)/\sqrt{3}$  & $2\pi/3$        \\
       & 4    &$(-1,1,1)/\sqrt{3}$ & $-2\pi/3$       \\
       & 5    &$(-1,1,1)/\sqrt{3}$ & $2\pi/3$        \\
       & 6    &$(-1,-1,1)/\sqrt{3}$& $-2\pi/3$       \\
       & 7    &$(-1,-1,1)/\sqrt{3}$& $2\pi/3$        \\
       & 8    &$(1,-1,1)/\sqrt{3}$ & $-2\pi/3$       \\
       & 9    &$(1,-1,1)/\sqrt{3}$ & $2\pi/3$        \\ \hline
 \multirow{6}{*}{$6C_4$}& 10   &$(1,0,0)$  & $-\pi /2$       \\
       & 11   &$(1,0,0)$  & $\pi /2$        \\
       & 12   &$(0,1,0)$  & $-\pi /2$       \\
       & 13   &$(0,1,0)$  & $\pi /2$        \\
       & 14   &$(0,0,1)$  &  $-\pi /2$      \\
       & 15   &$(0,0,1)$  &  $\pi /2$       \\ \hline
 \multirow{6}{*}{$6C'_2$}& 16   &$(0,1,1)/\sqrt{2}$  &  $-\pi$         \\
       & 17   &$(0,-1,1)/\sqrt{2}$ &  $-\pi$         \\
       & 18   &$(1,1,0)/\sqrt{2}$  &  $-\pi$         \\
       & 19   &$(1,-1,0)/\sqrt{2}$ &  $-\pi$         \\
       & 20   &$(1,0,1)/\sqrt{2}$  &  $-\pi$         \\
       & 21   &$(-1,0,1)/\sqrt{2}$ &  $-\pi$         \\ \hline
 \multirow{3}{*}{$3C_2$}& 22   &$(1,0,0)$  &  $-\pi$         \\
       & 23   &$(0,1,0)$  &  $-\pi$         \\
       & 24   &$(0,0,1)$  &  $-\pi$         \\
\hline
\end{tabular}
\end{center}
\caption{The elements of the cubic group, parameterized by the
  rotation axis $\vn^{(i)}$ and rotation angle $\omega_i$, divided into
    the different conjugacy classes.} 
\label{Table:RotationsOfO}
\end{table}
\begin{table}[t]
\begin{center}
\begin{tabular}{|c|c|c|c|c|c|c|c|c|}\cline{1-4} \cline{6-9}
 Class   &$i$  &$\vn^{(i)}$  & $\omega_i$    && Class & $i$   &
 $\vn^{(i)}$  & $\omega_i$ \\ \cline{1-4} \cline{6-9}
$I$      & 1   & any       &  $0$        & & \multirow{8}{*}{$8C_3$}& 28    & $(1,1,1)/\sqrt{3}$  & $4\pi/3$ \\ \cline{1-4}
\multirow{6}{*}{$6C_4$}   & 2   &$(1,0,0)$  &  $\pi$      &       && 29    & $(-1,1,1)/\sqrt{3}$ & $4\pi/3$ \\
         & 3   &$(0,1,0)$  &  $\pi$      &       && 30    & $(-1,-1,1)/\sqrt{3}$& $4\pi/3$ \\
         & 4   &$(0,0,1)$  &  $\pi$      &       && 31    & $(1,-1,1)/\sqrt{3}$ & $4\pi/3$ \\
         & 5   &$(1,0,0)$  &  $-\pi$     &       && 32    & $(1,1,1)/\sqrt{3}$  & $-4\pi/3$ \\
         & 6   &$(0,1,0)$  &  $-\pi$     &       && 33    & $(-1,1,1)/\sqrt{3}$ & $-4\pi/3$ \\
         & 7   &$(0,0,1)$  &  $-\pi$     &       && 34    & $(-1,-1,1)/\sqrt{3}$& $-4\pi/3$ \\ \cline{1-4}
\multirow{6}{*}{$6C'_8$}  & 8   &$(1,0,0)$  &  $\pi/2$    &       && 35    & $(1,-1,1)/\sqrt{3}$ & $-4\pi/3$ \\ \cline{6-9}
         & 9   &$(0,1,0)$  &  $\pi/2$    &&\multirow{12}{*}{$12C'_4$}& 36    & $(0,1,1)/\sqrt{2}$  & $\pi$    \\
         & 10  &$(0,0,1)$  &  $\pi/2$    &       && 37    & $(0,-1,1)/\sqrt{2}$ & $\pi$    \\
         & 11  &$(1,0,0)$  &  $-\pi/2$   &       && 38    & $(1,1,0)/\sqrt{2}$  & $\pi$    \\
         & 12  &$(0,1,0)$  &  $-\pi/2$   &       && 39    & $(1,-1,0)/\sqrt{2}$ & $\pi$    \\
         & 13  &$(0,0,1)$  &  $-\pi/2$   &       && 40    & $(1,0,1)/\sqrt{2}$  & $\pi$    \\ \cline{1-4}
\multirow{6}{*}{$6C_8$}   & 14  &$(1,0,0)$  &  $3\pi/2$   &       && 41    & $(-1,0,1)/\sqrt{2}$ & $\pi$    \\
         & 15  &$(0,1,0)$  &  $3\pi/2$   &       && 42    & $(0,1,1)/\sqrt{2}$  & $-\pi$   \\
         & 16  &$(0,0,1)$  &  $3\pi/2$   &       && 43    & $(0,-1,1)/\sqrt{2}$ & $-\pi$   \\
         & 17  &$(1,0,0)$  &  $-3\pi/2$  &       && 44    & $(1,1,0)/\sqrt{2}$  & $-\pi$   \\
         & 18  &$(0,1,0)$  &  $-3\pi/2$  &       && 45    & $(1,-1,0)/\sqrt{2}$ & $-\pi$   \\
         & 19  &$(0,0,1)$  &  $-3\pi/2$  &       && 46    & $(1,0,1)/\sqrt{2}$  & $-\pi$   \\ \cline{1-4}
\multirow{8}{*}{$8C_6$}   & 20  &$(1,1,1)/\sqrt{3}$  &  $2\pi/3$   &       && 47    & $(-1,0,1)/\sqrt{2}$  & $-\pi$   \\ \cline{6-9}
         & 21  &$(-1,1,1)/\sqrt{3}$ &  $2\pi/3$   &&  $J$  & 48    & any  & $2\pi$   \\ \cline{6-9}
         & 22  &$(-1,-1,1)/\sqrt{3}$&  $2\pi/3$   &  \multicolumn{4}{c}{$ $} \\
         & 23  &$(1,-1,1)/\sqrt{3}$ &  $2\pi/3$   &  \multicolumn{4}{c}{$ $} \\
         & 24  &$(1,1,1)/\sqrt{3}$  &  $-2\pi/3$  &  \multicolumn{4}{c}{$ $} \\
         & 25  &$(-1,1,1)/\sqrt{3}$ &  $-2\pi/3$  &  \multicolumn{4}{c}{$ $} \\
         & 26  &$(-1,-1,1)/\sqrt{3}$&  $-2\pi/3$  &  \multicolumn{4}{c}{$ $} \\
         & 27  &$(1,-1,1)/\sqrt{3}$ &  $-2\pi/3$  &  \multicolumn{4}{c}{$ $} \\ \cline{1-4}
\end{tabular}
\end{center}
\caption{The elements of the double cover group of the cubic group,
  parameterized by the 
  rotation axis $\vn^{(i)}$ and rotation angle $\omega_i$, divided into
    the different conjugacy classes.}
\label{Table:RotationsOf2O}
\end{table}

The full symmetry group includes space inversions $I$, which commute
with the elements of $O$ and $^2O$. Choosing
$T(I)=-1$,\footnote{Alternatively, we could have chosen $T(I)=1$. Both
  choices are consistent with $I^2=1$.} where $T(I)$ denotes
an element of any matrix representation of $I$, the elements 
of $O$ and $^2O$ combined with $I$ form the product groups
$O_h=O\otimes\{1,-1\}$ and $^2O_h={}^2O\otimes\{1,-1\}$, respectively.
Irreducible matrix representations of $O_h$ and $^2O_h$
have been given, for example, in~\cite{Bernard:2008ax}.

In the CM frame moving with velocity $\vv = \vP/W$, $\vP =
(2\pi/L)\, \vd$ in the laboratory frame, the symmetry
group reduces to 
certain subgroups of $O_h$ and $^2O_h$, hereafter referred to as the
little groups. In the case of unequal masses, the little group consists
of elements $S_i = \{R_i,I R_i\}\in O_h$ and $^2O_h$, respectively,
which obey 
\eq
S_i\,\vd=\vd\, .
\en
In the case of equal masses, the system is symmetric under $\vd
\rightarrow -\vd$, and the little group consists of elements $S_i$,
which obey
\eq
S_i\, \vd =\pm \vd\, .
\en
In Table~\ref{Table:i} we list the elements $R_i \in O$ and $^2O$
that satisfy the condition $R_i\, \vd=\vd$ and $R_i\, \vd=-\vd$ for
our three choices of $\vd$, together with the corresponding little
groups. With $I \vd = -\vd$, the action of the group elements $S_i$ on
$\vd$ is now fully defined.

\begin{table}[t]
\begin{center}
\begin{tabular}{|c|c|c|c|c|}
\hline
Group & $\vd$ & Little Group & $R_i\,\vd=\vd$ & $R_i\,\vd=-\vd$ \\
\hline
\multirow{3}{*}{$O_h$} & $(0,0,1)$ & $C_{4v}$& $\{R_i|i=1,14,15,24\}$& $\{R_i|i=18,19,22,23\}$\\
& $(1,1,0)$ & $C_{2v}$& $\{R_i|i=1,18\}$& $\{R_i|i=19,24\}$\\
& $(1,1,1)$ & $C_{3v}$& $\{R_i|i=1,2,3\}$& $\{R_i|i=17,19,21\}$\\
\hline
\multirow{3}{*}{$^2O_h$} & $(0,0,1)$ & $^2C_{4v}$& $\{R_i|i=1,4,7,10,13,16,19,48\}$& 
$\{R_i|i=2,3,5,6,38,39,44,45\}$\\
& $(1,1,0)$ & $^2C_{2v}$& $\{R_i|i=1,38,44,48\}$& $\{R_i|i=4,7,39,45\}$\\
& $(1,1,1)$ & $^2C_{3v}$& $\{R_i|i=1,20,24,28,32,48\}$& $\{R_i|i=37,39,41,43,45,47\}$\\
\hline
\end{tabular}
\end{center}
\caption{Rotations $R_i$ that obey the condition
$R_i\,\vd=\pm \vd$.}
\label{Table:i}
\end{table}

As we shall see, several of the irreducible representations of the
little groups $C_{4v}$, $^2C_{4v}$, $^2C_{2v}$, $C_{3v}$ and
$^2C_{3v}$ in Table~\ref{Table:i}, namely $E$, $G_1$ and $G_2$, are 
two-dimensional. Two-dimensional representations $G_1$ and $G_2$ can
be built from the matrices 
\eq
 (Y_i)_{\alpha\beta}=\left(e^{-\frac{i}{2}\,\vn^{(i)}
\mathbold{\sigma}\,\omega_i}\right)_{\alpha\beta}=\delta_{\alpha\beta}\,
\cos\frac{\omega_i}{2}  
-i\,\left(\vn^{(i)}\mathbold{\sigma}\right)_{\alpha\beta}\,\sin\frac{\omega_i}{2}\,,
\quad \alpha,\beta=1,2 
\label{Eqn:R_for_2O}
\en
with $i=1,\cdots,48$. 

For the two-dimensional representation $E$ in
the bosonic case, it is convenient to introduce the matrices
\begin{equation}
\begin{split}
&X_1=\mathbbm{1}\, ,\quad
X_2=-\frac{1}{2}\,\mathbbm{1}+i\frac{\sqrt{3}}{2}\,\sigma_2\, ,\quad
X_3=-\frac{1}{2}\,\mathbbm{1}-i\frac{\sqrt{3}}{2}\,\sigma_2\, ,\quad
X_4=-\frac{1}{2}\,\sigma_3-\frac{\sqrt{3}}{2}\,\sigma_1\, ,\quad
\\%[2mm]
&X_5=\sigma_3\, ,\quad
X_6=-\frac{1}{2}\,\sigma_3+\frac{\sqrt{3}}{2}\,\sigma_1\,  ,\quad 
X_7= i\frac{1}{\sqrt{2}}\,(\sigma_1+\sigma_2)\,  ,\quad X_8=
\frac{1}{\sqrt{2}}\,(\sigma_1-\sigma_2)\,  .
\end{split}
\end{equation}

\begin{table}[t!]
\renewcommand{\arraystretch}{0.95}
\begin{center}
\begin{tabular}{|c|c|c|c|c|c|}\hline
$\{S_i\}$     &  $R_{1}$    & $\{R_{14}, R_{15}\}$  &  $\{IR_{18}, IR_{19}\}$    &   $\{IR_{22}, IR_{23}\}$    &   $R_{24}$  \\
\hline\hline
\backslashbox{$\Gamma$\kern-0em}{Class} & $I$ & $2C_4$ & $2 IC_2^\prime$ & $2 IC_2$   & $C_2$ \\\hline
$A_1$     &  1  &    1   &    1   &       1         &   1     \\
$A_2$     &  1  &    1   &   -1   &      -1         &   1     \\
$B_1$     &  1  &   -1   &   -1   &       1         &   1     \\
$B_2$     &  1  &   -1   &    1   &      -1         &   1     \\
$E $     &  2  &    0   &    0   &       0         &   -2     \\
\hline\hline
$E$ & $X_1$ & $\{-X_7,X_7\}$ & $\{X_5,-X_5\}$ & $\{X_8,-X_8\}$ &   $-X_1$      \\
\hline
\end{tabular}
\end{center}
\caption{Character table of the little group $C_{4v}$ for integer
  spin. The top row shows the elements of $C_{4v}$ divided into
  conjugacy classes. The bottom row shows the corresponding matrices
  for the two-dimensional irreducible
  representation $E$. The rows in between list the
  characters of the various irreducible representations $\Gamma$ of
  $C_{4v}$.}  
\label{Table:Characters_Bos_DiffM_001}
%\end{table}
%
%\begin{sidewaystable}
%\begin{table}
%\begin{table}
%\renewcommand{\arraystretch}{0.8}
%\vspace*{0.5cm}
\begin{center}\fontsize{10pt}{14pt}\selectfont
\begin{tabular}{|c|c|c|c|c|c|c|c|}\hline\fontsize{10pt}{12pt}\selectfont
$\{S_i\}$   &  $R_1$  &   $\{R_{4}, R_{7}\}$    & $\{R_{10}, R_{13}\}$ & $\{R_{16}, R_{19}\}$  &  $\{IR_{2}, IR_{3},IR_{5}, IR_{6}\}$    & $\{IR_{38}, IR_{39}, IR_{44}, IR_{45}\}$     &  $R_{48}$   \\
\hline\hline
\backslashbox{$\Gamma$\kern-1em}{Class}  & $I$ & $2C_4$ & $2C_8^\prime$ & $2C_8$ & $4 IC_4$ &  $4 IC_4^\prime$  & $J$ \\\hline
$A_1$    &  1  &    1   &    1        &      1      &   1    &   1    &   1    \\
$A_2$    &  1  &    1   &    1        &      1      &  -1    &  -1    &   1    \\
$B_1$    &  1  &    1   &   -1        &     -1      &   1    &  -1    &   1    \\
$B_2$    &  1  &    1   &   -1        &     -1      &  -1    &   1    &   1    \\
$E $    &  2  &   -2   &    0        &      0      &   0    &   0    &   2    \\
$G_1$    &  2  &    0   & $\sqrt{2}$  & $-\sqrt{2}$ &   0    &   0    &  -2    \\
$G_2$    &  2  &    0   & $-\sqrt{2}$ & $\sqrt{2}$  &   0    &   0    &  -2    \\
\hline\hline
 $E$    &  $\mathbbm{1}$  &  $\{-\mathbbm{1}, -\mathbbm{1}\}$    & $\{i\sigma_3, -i\sigma_3\}$  &  $\{-i\sigma_3, i\sigma_3\}$ &  $\{\sigma_1, -\sigma_1, \sigma_1, -\sigma_1\}$   &  $\{-\sigma_2, \sigma_2, -\sigma_2, \sigma_2\}$    &  $\mathbbm{1}$    \\
 $G_1$   &  $Y_1$  &   $\{Y_{4}, Y_{7}\}$    & $\{Y_{10}, Y_{13}\}$ & $\{Y_{16}, Y_{19}\}$  &  $\{-Y_{2}, -Y_{3},-Y_{5}, -Y_{6}\}$    & $\{-Y_{38}, -Y_{39}, -Y_{44}, -Y_{45}\}$     &  $Y_{48}$ 
 \\
 $G_2$   &  $Y_1$  &   $\{Y_{4}, Y_{7}\}$    & $\{-Y_{10}, -Y_{13}\}$
 & $\{-Y_{16}, -Y_{19}\}$  &  $\{Y_{2}, Y_{3},Y_{5}, Y_{6}\}$    &
 $\{-Y_{38}, -Y_{39}, -Y_{44}, -Y_{45}\}$     &  $Y_{48}$   \\
\hline
\end{tabular}
\end{center}
\caption{The same as Table~\ref{Table:Characters_Bos_DiffM_001} for
    the little group $^2C_{4v}$ and half-integer spin, together
    with the matrices of the two-dimensional irreducible
  representations $E$, $G_1$ and $G_2$.}
\label{Table:Characters_Ferm_DiffM_001}
%\end{table}
%\end{sidewaystable}
\end{table}

\begin{table}[t]
\renewcommand{\arraystretch}{0.9}
\begin{center}
\begin{tabular}{|r|c|c|c|c|}\hline
$ S_i$   &  $R_1$  &  $R_{18}$    &   $IR_{19}$   &   $IR_{24}$  \\
\hline\hline
 \backslashbox{$\Gamma$\kern-0em}{Class}& $I$ & $C_2^\prime$ & $IC_2^\prime$ &  $IC_2$  \\\hline
$A_1$    &  1  &    1   &    1   &       1           \\
$A_2$    &  1  &    1   &   -1   &      -1           \\
$B_1$    &  1  &   -1   &    1   &      -1           \\
$B_2$    &  1  &   -1   &   -1   &       1           \\
\hline
\end{tabular}
\end{center}
\caption{The same as Table~\ref{Table:Characters_Bos_DiffM_001} for
    the little group $C_{2v}$ and integer spin.}
\label{Table:Characters_Bos_DiffM_110}
%\end{table}
%
%\begin{table}[t]
\renewcommand{\arraystretch}{0.9}
\vspace*{0.25cm}
\begin{center}
\begin{tabular}{|c|c|c|c|c|c|}\hline
$\{S_i\}$    &  $R_{1}$  &   $\{R_{38}, R_{44}\}$    &  $\{IR_{4}, IR_{7}\}$     &   $\{IR_{39}, IR_{45}\}$     &    $R_{48}$       \\
\hline\hline
 \backslashbox{$\Gamma$\kern-0em}{Class}& $I$ & $2C_4^\prime$ & $4 IC_4$ & $4 IC_4^\prime$ & $J$ \\\hline
$A_1$    &  1  &    1   &    1   &       1 &       1          \\
$A_2$    &  1  &    1   &   -1   &      -1 &       1          \\
$B_1$    &  1  &   -1   &   -1   &       1 &       1          \\
$B_2$    &  1  &   -1   &    1   &      -1 &       1          \\
$G_1$    &  2  &    0   &    0   &       0 &      -2          \\
\hline\hline
$G_1$ & $Y_1$ & $\{Y_{38}, Y_{44}\}$ & $\{-Y_4, -Y_7\}$ & $\{-Y_{39},-Y_{45}\}$ & $Y_{48}$ \\
\hline
\end{tabular}
\end{center}
\caption{The same as Table~\ref{Table:Characters_Bos_DiffM_001} for
    the little group $^2C_{2v}$ and half-integer spin, together
    with the matrices of the two-dimensional irreducible
  representation $G_1$.}
\label{Table:Characters_Ferm_DiffM_110}
%\end{table}
%
%\begin{table}[t]
\renewcommand{\arraystretch}{0.9}
\vspace*{0.25cm}
\begin{center}
\begin{tabular}{|c|c|c|c|}\hline
$\{S_i\}$   &  $R_1$  &  $\{R_2,R_3\}$    &   $\{IR_{17}, IR_{19}, IR_{21}\}$  \\
\hline\hline
 \backslashbox{$\Gamma$\kern-0em}{Class}& $I$ & $2C_3$ & $3 IC_2^\prime$ \\\hline
$A_1$    &  1  &    1   &    1       \\
$A_2$    &  1  &    1   &   -1        \\
$E$     &  2  &   -1   &    0         \\
\hline
 $E$ & $X_1$ & $\{X_2,X_3\}$ & $\{-X_4,-X_5,-X_6\}$\\
\hline
\end{tabular}
\end{center}
\caption{The same as Table~\ref{Table:Characters_Bos_DiffM_001} for
    the little group $C_{3v}$ and integer spin, together
    with the matrices of the two-dimensional irreducible
  representation $E$.}
\label{Table:Characters_Bos_DiffM_111}
\end{table}

\begin{table}[t]
\renewcommand{\arraystretch}{1.5}
\vspace*{1cm}
\begin{center}
\begin{tabular}{|c|c|c|c|c|c|c|}\hline
$\{S_i\}$    &  $R_{1}$  &   $\{R_{20},R_{24}\}$    &   $\{R_{28},R_{32}\}$    &  $\{IR_{37},I R_{45}, IR_{47}\}$      & $\{IR_{39},IR_{41}, IR_{43}\}$   &   $R_{48}$       \\
\hline\hline
\backslashbox{$\Gamma$\kern-0em}{Class} & $I$ & $2C_6$ & $2C_3$ & $3 IC_4$ & $3 IC_4^\prime$ & $J$\\\hline
$A_1$    &  1  &    1   &    1   &    1    &    1   &    1      \\
$A_2$    &  1  &    1   &    1   &   -1    &   -1   &    1      \\ 
$K_1$    &  1  &   -1   &    1   &   $i$   &  $-i$  &   -1      \\
$K_2$    &  1  &   -1   &    1   &  $-i$   &   $i$  &   -1      \\
$E $    &  2  &   -1   &   -1   &    0    &    0   &    2      \\
$G_1$    &  2  &    1   &   -1   &    0    &    0   &   -2      \\     
\hline\hline
$E$    &  $X_{1}$  &  $\{X_{3}, X_{2}\}$    &  $\{X_{2}, X_{3}\}$   &  $\{-X_4, -X_5, -X_6\}$      &  $\{-X_5, -X_6, -X_4\}$    &   $X_{1}$       \\
%\hline
$G_1$ & $Y_1$ &  $\{Y_{20}, Y_{24}\}$  & $\{Y_{28}, Y_{32}\}$ &
  $\{-Y_{37}, -Y_{45}, -Y_{47}\}$ & $\{-Y_{39}, -Y_{41}, -Y_{43}\}$ & $Y_{48}$ \\ 
\hline
\end{tabular}
\end{center}
\caption{The same as Table~\ref{Table:Characters_Bos_DiffM_001} for
    the little group $^2C_{3v}$ and half-integer spin, together
    with the matrices of the two-dimensional irreducible
  representation $E$ and $G_1$.}
\label{Table:Characters_Ferm_DiffM_111}
\end{table}

In Tables~\ref{Table:Characters_Bos_DiffM_001},
\ref{Table:Characters_Ferm_DiffM_001},
\ref{Table:Characters_Bos_DiffM_110},
\ref{Table:Characters_Ferm_DiffM_110},
\ref{Table:Characters_Bos_DiffM_111},
\ref{Table:Characters_Ferm_DiffM_111},
we list the elements $\{S_i\}$ of the little groups $C_{4v}$, $^2C_{4v}$,
$^2C_{2v}$, $C_{3v}$ and $^2C_{3v}$, broken into the various conjugacy
classes, the irreducible representations $\Gamma$ and characters
$\chi(\Gamma)$ of the little groups. Note that the rotation matrices
$R_i$ are specified by the rotation axes $\vn^{(i)}$ and angles
$\omega_i$ given in Tables~\ref{Table:RotationsOfO} and
\ref{Table:RotationsOf2O} for the bosonic and fermionic case,
respectively. 
In the case of the two-dimensional
representations, we additionally list the matrices $X_i$, $Y_i$
corresponding to the respective group elements. It is straightforward
to check that these matrices obey the group multiplication laws. Later
on we will need the whole information communicated in the tables for
the construction of basis 
vectors and operators that transform according to the individual irreducible
representations.

The results hold for the general case of unequal masses. For equal
masses (integer 
spin) the representations are merely `doubled'. Let us explain this by
giving a specific example. Consider the representation $A_1$ in
Table~\ref{Table:Characters_Bos_DiffM_001}. For equal masses two
representations emerge: $A_1^+$ corresponding to $T(S)=1$ for $S=R_1$,
$R_{14}$, $R_{15}$, $R_{24}$, $R_{18}$, 
$R_{19}$, $R_{22}$, $R_{23}$, $IR_1$, $IR_{14}$, $IR_{15}$,
$IR_{24}$, $IR_{18}$, 
$IR_{19}$, $IR_{22}$,
$IR_{23}$, and $A_1^-$ corresponding to  $T(S)=1$ for $S=R_1$,
$R_{14}$, $R_{15}$, $R_{24}$, $IR_{18}$, $IR_{19}$, $IR_{22}$, $IR_{23}$ and
$T(S)=-1$ for $S=IR_1$, $IR_{14}$, $IR_{15}$, $IR_{24}$, $R_{18}$, $R_{19}$,
$R_{22}$, $R_{23}$. All other representations are `doubled' in a similar
manner. 

\section{Phase shifts}
\label{sec5}

From now on we shall drop the superscript $\vDelta$ from the 
matrix 
$M$. Following~\cite{Bernard:2008ax}, the scattering phase shifts
$\delta$ are obtained from the determinant equation
\begin{equation}
\det\,\left(M_{lm,l^\prime m^\prime}- \delta_{l l^\prime} \delta_{m
  m^\prime} \cot \delta_l \right) = 0
\label{boson}
\end{equation}
for meson resonances and 
\begin{equation}
\det\,\left(M_{Jl\mu,J^\prime l^\prime \mu^\prime}- \delta_{J
  J^\prime}\delta_{l l^\prime} \delta_{\mu 
  \mu^\prime} \cot \delta_{Jl} \right) = 0
\label{baryon}
\end{equation}
for baryon resonances. Equations (\ref{boson}) and (\ref{baryon})
relate the phases $\delta$ in the infinite volume to the energy levels
of the lattice Hamiltonian in a finite cubic box.

\subsection{Reduction of the phase shift formulae}

\label{sec:cubic}

In the infinite volume, the basis vectors of an irreducible representation
$D^l$ of the rotation group of total angular momentum $l$
are given by $|lm\rangle$ with $M_{lm,l^\prime m^\prime}=\langle l
m|\hat{M}|l^\prime m^\prime\rangle$ . In case of half-integer spin, the basis
vectors are given by
\begin{equation}
|Jl\mu\rangle=\sum_{m,\sigma}|lm,\frac{1}{2}\sigma\rangle
\langle l m,\frac{1}{2}\sigma|J\mu\rangle 
\end{equation}
with
\begin{equation}
M_{Jl\mu,J^\prime l^\prime \mu^\prime} = \langle Jl\mu|\hat{M}|J^\prime
  l^\prime \mu^\prime\rangle \,,
\end{equation}
where $\mu=-J,\ldots, +J$ and $l=J\pm\frac{1}{2}$. The vectors
$|lm\rangle$ and $|Jl\mu\rangle$ are parity eigenstates with parity
$(-1)^l$.  

In the case of the moving frame the basis vectors of an irreducible
representation $\Gamma$ can be written as
\begin{equation}
|\Gamma\alpha l n\rangle=\sum_mc^{\Gamma \alpha n}_{lm}|lm\rangle 
\end{equation}
for integer spin and
\begin{equation}
|\Gamma\alpha J l n\rangle=\sum_\mu c^{\Gamma \alpha n}_{Jl\mu}|Jl\mu\rangle 
\end{equation}
for half-integer spin, where $\alpha$ runs from $1$ to the
dimension of $\Gamma$, and $n$ runs from $1$ to $N(\Gamma,l)$, the
number of occurrences of the irreducible representation $\Gamma$ in
$D^l$. The basis vectors of the various frames and representations are
given in Tables~\ref{Table:Basis_Bos_DiffM_001},
\ref{Table:Basis_Ferm_DiffM_001},
\ref{Table:Basis_Bos_DiffM_110},
\ref{Table:Basis_Ferm_DiffM_110},
\ref{Table:Basis_Bos_DiffM_111},
\ref{Table:Basis_Ferm_DiffM_111} for $l=0, 1$ and $2$. 
The coefficients $c^{\Gamma \alpha n}_{lm}$ and $c^{\Gamma \alpha n}_{Jl\mu}$ 
can be directly read off from these tables.

\begin{table}
\begin{center}
\begin{tabular}{|c|r|r|c|}\hline
$\Gamma$ & $l$ & $\alpha$ & Basis vectors \\
\hline\hline
$A_1$ & 0 &  & $|0, 0\rangle$\\
$A_1$ & 1 &  & $|1, 0\rangle$\\
\multirow{2}{*}{$E$}   & \multirow{2}{*}{1} & 1 & $\frac{1}{2}(1-i)|1, -1\rangle-\frac{1}{2}(1+i)|1, 1\rangle$\\
      &   & 2 & $-\frac{1}{\sqrt{2}}i|1, -1\rangle+\frac{1}{\sqrt{2}}|1, 1\rangle$\\
$A_1$ & 2 &  & $|2, 0\rangle$\\
$B_1$ & 2 &  & $\frac{1}{\sqrt{2}}(|2, -2\rangle+|2, 2\rangle)$\\
$B_2$ & 2 &  & $\frac{1}{\sqrt{2}}(|2, -2\rangle-|2, 2\rangle)$\\
\multirow{2}{*}{$E$}   & \multirow{2}{*}{2} & 1 & $\frac{1}{\sqrt{2}}|2, -1\rangle-\frac{1}{\sqrt{2}}i|2, 1\rangle$\\
      &   & 2 & $\frac{1}{2}(1-i)|2, -1\rangle+\frac{1}{2}(1+i)|2, 1\rangle$\\
\hline
\end{tabular}
\end{center}
\caption{The basis vectors
 of the irreducible representations $\Gamma$ of the little group
 $C_{4v}$ and integer spin. $\alpha$ labels the components of
 the basis vectors of the two-dimensional representation $E$.} 
\label{Table:Basis_Bos_DiffM_001}
%\end{table}
%
%\begin{table}
%\renewcommand{\arraystretch}{1.5}
\vspace*{0.25cm}
\begin{center}
\begin{tabular}{|c|c|c|c|c|}\hline
$\Gamma$ & $J$ & $l$ & $\alpha$ & Basis vectors  \\
\hline\hline
\multirow{2}{*}{$G_1$} & \multirow{2}{*}{$\frac{1}{2}$} & \multirow{2}{*}{0} & 1 & $|\frac{1}{2}, \frac{1}{2}\rangle$\\
      &               &   & 2 & $-|\frac{1}{2}, -\frac{1}{2}\rangle$\\
\multirow{2}{*}{$G_1$} & \multirow{2}{*}{$\frac{1}{2}$} & \multirow{2}{*}{1} & 1 & $|\frac{1}{2}, \frac{1}{2}\rangle$\\
      &               &   & 2 & $|\frac{1}{2}, -\frac{1}{2}\rangle$\\
\multirow{2}{*}{$G_1$} & \multirow{2}{*}{$\frac{3}{2}$} & \multirow{2}{*}{1} & 1 & $|\frac{3}{2}, \frac{1}{2}\rangle$\\
      &               &   & 2 & $-|\frac{3}{2}, -\frac{1}{2}\rangle$\\
\multirow{2}{*}{$G_2$} & \multirow{2}{*}{$\frac{3}{2}$} & \multirow{2}{*}{1} & 1 & $|\frac{3}{2}, -\frac{3}{2}\rangle$\\
      &               &   & 2 & $|\frac{3}{2}, \frac{3}{2}\rangle$\\
\multirow{2}{*}{$G_1$} & \multirow{2}{*}{$\frac{3}{2}$} & \multirow{2}{*}{2} & 1 & $|\frac{3}{2}, \frac{1}{2}\rangle$\\
      &               &   & 2 & $|\frac{3}{2}, -\frac{1}{2}\rangle$\\
\multirow{2}{*}{$G_2$} & \multirow{2}{*}{$\frac{3}{2}$} &\multirow{2}{*}{2} & 1 & $|\frac{3}{2}, -\frac{3}{2}\rangle$\\
      &               &   & 2 & $-|\frac{3}{2}, \frac{3}{2}\rangle$\\
\hline
\end{tabular}
\end{center}
\caption{The basis vectors
 of the irreducible representations $\Gamma$ of the little group
 $^2C_{4v}$ and half-integer spin. $\alpha$ labels the components of
 the basis vectors of the two-dimensional representations $G_1$ and $G_2$.} 
\label{Table:Basis_Ferm_DiffM_001}
\end{table}

\begin{table}
\begin{center}
\begin{tabular}{|c|c|c|}\hline
$\Gamma$ &$l$ & Basis vectors \\
\hline\hline
$A_1$ & 0 & $|0,0\rangle$\\
$A_1$ & 1 & $\frac{1}{\sqrt{2}}(-i|1,-1\rangle+|1,1\rangle)$\\
$B_1$ & 1 & $|1,0\rangle$\\
$B_2$ & 1 & $\frac{1}{\sqrt{2}}(i|1,-1\rangle+|1,1\rangle)$\\
$A_1$ & 2 & $|2,0\rangle$\\
$A_1$ & 2 & $\frac{1}{\sqrt{2}}(|2,-2\rangle-|2,2\rangle)$\\
$A_2$ & 2 & $\frac{1}{\sqrt{2}}(|2,-1\rangle-i|2,1\rangle)$\\
$B_1$ & 2 & $\frac{1}{\sqrt{2}}(|2,-1\rangle+i|2,1\rangle)$\\
$B_2$ & 2 & $\frac{1}{\sqrt{2}}(|2,-2\rangle+|2,2\rangle)$\\
\hline
\end{tabular}
\end{center}
\caption{The basis vectors
 of the irreducible representations $\Gamma$ of the little group
 $C_{2v}$ and integer spin.} 
\label{Table:Basis_Bos_DiffM_110}
%\end{table}
%
%\begin{table}
%\renewcommand{\arraystretch}{1.5}
\vspace*{0.25cm}
\begin{center}
\begin{tabular}{|l|r|c|c|c|}\hline
$\Gamma$ & $J$ & $l$ & $\alpha$ & Basis vectors  \\
\hline\hline
\multirow{2}{*}{$G_1$} & \multirow{2}{*}{$\frac{1}{2}$} & \multirow{2}{*}{0} & 1 & $|\frac{1}{2},-\frac{1}{2}\rangle$\\
 &  &  & 2 & $-i|\frac{1}{2},\frac{1}{2}\rangle$\\
\multirow{2}{*}{$G_1$} & \multirow{2}{*}{$\frac{1}{2}$} & \multirow{2}{*}{1} & 1 & $|\frac{1}{2},\frac{1}{2}\rangle$\\
 &  &  & 2 & $|\frac{1}{2},-\frac{1}{2}\rangle$\\
\multirow{2}{*}{$G_1$} & \multirow{2}{*}{$\frac{3}{2}$} & \multirow{2}{*}{1} & 1 & $|\frac{3}{2},-\frac{3}{2}\rangle$\\
 &  &  & 2 & $|\frac{3}{2},\frac{3}{2}\rangle$\\
\multirow{2}{*}{$G_1$} & \multirow{2}{*}{$\frac{3}{2}$} & \multirow{2}{*}{1} & 1 & $-|\frac{3}{2},\frac{1}{2}\rangle$\\
 &  &  & 2 & $|\frac{3}{2},-\frac{1}{2}\rangle$\\
\multirow{2}{*}{$G_1$} & \multirow{2}{*}{$\frac{3}{2}$} & \multirow{2}{*}{2} & 1 & $i|\frac{3}{2},\frac{3}{2}\rangle$\\
 &  &  & 2 & $|\frac{3}{2},-\frac{3}{2}\rangle$\\
\multirow{2}{*}{$G_1$} & \multirow{2}{*}{$\frac{3}{2}$} & \multirow{2}{*}{2} & 1 & $|\frac{3}{2},-\frac{1}{2}\rangle$\\
 &  &  & 2 & $i|\frac{3}{2},\frac{1}{2}\rangle$\\
\hline
\end{tabular}
\end{center}
\caption{The basis vectors
 of the irreducible representations $\Gamma$ of the little group
 $^2C_{2v}$ and half-integer spin. $\alpha$ labels the components of
 the basis vectors of the two-dimensional representation $G_1$.}  
\label{Table:Basis_Ferm_DiffM_110}
\end{table}

\begin{table}
\begin{center}
\begin{tabular}{|l|r|r|c|}\hline
$\Gamma$ & $l$ & $\alpha$ & Basis vectors \\
\hline\hline
$A_1$ & 0 &  & $|0, 0\rangle$\\
$A_1$ & 1 &  & $-\frac{1}{\sqrt{3}}i|1, -1\rangle-\frac{1}{\sqrt{6}}(1+i)|1, 0\rangle+\frac{1}{\sqrt{3}}|1, 1\rangle$\\
\multirow{2}{*}{$E$}   & \multirow{2}{*}{1} & 1 & $\frac{1}{\sqrt{2}}i|1, -1\rangle+\frac{1}{\sqrt{2}}|1, 1\rangle$\\
      &   & 2 & $-\frac{1}{\sqrt{6}}|1, -1\rangle+\frac{1}{\sqrt{3}}(1-i)|1, 0\rangle-\frac{1}{\sqrt{6}}i|1, 1\rangle$\\
$A_1$ & 2 &  & $\frac{1}{\sqrt{6}}|2, -2\rangle+\frac{1}{\sqrt{6}}(1-i)|2, -1\rangle+\frac{1}{\sqrt{6}}(1+i)|2, 1\rangle-\frac{1}{\sqrt{6}}|2, 2\rangle$\\
\multirow{2}{*}{$E$}   & \multirow{2}{*}{2} & 1 & $\frac{1}{\sqrt{2}}|2, -2\rangle+\frac{1}{\sqrt{2}}|2, 2\rangle$\\
      &   & 2 & $-|2, 0\rangle$\\
\multirow{2}{*}{$E$}   & \multirow{2}{*}{2} & 1 & $\frac{1}{\sqrt{2}}|2, -1\rangle-\frac{1}{\sqrt{2}}i|2, 1\rangle$\\
      &   & 2 & $-\frac{1}{\sqrt{6}}(1-i)|2, -2\rangle-\frac{1}{\sqrt{6}}i|2, -1\rangle+\frac{1}{\sqrt{6}}|2, 1\rangle+\frac{1}{\sqrt{6}}(1-i)|2, 2\rangle$\\
\hline
\end{tabular}
\end{center}
\caption{The basis vectors
 of the irreducible representations $\Gamma$ of the little group
 $C_{3v}$ and integer spin. $\alpha$ labels the components of
 the basis vectors of the two-dimensional representation $E$.} 
\label{Table:Basis_Bos_DiffM_111}
%\end{table}
%
%\begin{sidewaystable}
%\begin{table}
%\renewcommand{\arraystretch}{1.5}
\vspace*{0.25cm}
\begin{center}
\begin{tabular}{|c|c|c|c|c|}\hline
$\Gamma$ & $J$ & $l$ & $\alpha$ & Basis vectors  \\
\hline\hline
\multirow{2}{*}{$G_1$} & \multirow{2}{*}{$\frac{1}{2}$} & \multirow{2}{*}{0} & 1 & $\frac{\sqrt{6}}{3}|\frac{1}{2}, -\frac{1}{2}\rangle+\frac{1}{\sqrt{6}}(1-i)|\frac{1}{2}, \frac{1}{2}\rangle$\\
      &   &   & 2 & $-\frac{1}{\sqrt{6}}(1-i)|\frac{1}{2}, -\frac{1}{2}\rangle-\frac{\sqrt{6}}{3}i|\frac{1}{2}, \frac{1}{2}\rangle$\\
\multirow{2}{*}{$G_1$} & \multirow{2}{*}{$\frac{1}{2}$} & \multirow{2}{*}{1} & 1 & $|\frac{1}{2}, \frac{1}{2}\rangle$\\
      &   &   & 2 & $|\frac{1}{2}, -\frac{1}{2}\rangle$\\
\multirow{2}{*}{$G_1$} & \multirow{2}{*}{$\frac{3}{2}$} & \multirow{2}{*}{1} & 1 & $\frac{1}{\sqrt{6}}(1+i)|\frac{3}{2}, -\frac{3}{2}\rangle+\frac{1}{\sqrt{2}}|\frac{3}{2}, -\frac{1}{2}\rangle+\frac{1}{\sqrt{6}}i|\frac{3}{2}, \frac{3}{2}\rangle$\\
      &   &   & 2 & $-\frac{1}{\sqrt{6}}|\frac{3}{2}, -\frac{3}{2}\rangle-\frac{1}{\sqrt{2}}i|\frac{3}{2}, \frac{1}{2}\rangle+\frac{1}{\sqrt{6}}(1+i)|\frac{3}{2}, \frac{3}{2}\rangle$\\
$K_1$ & $\frac{3}{2}$ & 1 &  & $-(\frac{\sqrt{3}}{6}(1+i)+\frac{\sqrt{6}}{12}(1-i))|\frac{3}{2}, -\frac{3}{2}\rangle+\frac{1}{2}|\frac{3}{2}, -\frac{1}{2}\rangle+\frac{\sqrt{2}}{4}(1+i)|\frac{3}{2}, \frac{1}{2}\rangle+(\frac{\sqrt{6}}{6}-\frac{\sqrt{3}}{6}i)|\frac{3}{2}, \frac{3}{2}\rangle$\\
$K_2$ & $\frac{3}{2}$ & 1 &  & $-(\frac{\sqrt{3}}{6}(1+i)-\frac{\sqrt{6}}{12}(1-i))|\frac{3}{2}, -\frac{3}{2}\rangle+\frac{1}{2}|\frac{3}{2}, -\frac{1}{2}\rangle-\frac{\sqrt{2}}{4}(1+i)|\frac{3}{2}, \frac{1}{2}\rangle-(\frac{\sqrt{6}}{6}+\frac{\sqrt{3}}{6}i)|\frac{3}{2}, \frac{3}{2}\rangle$\\
\multirow{2}{*}{$G_1$} & \multirow{2}{*}{$\frac{3}{2}$} &\multirow{2}{*}{2} & 1 & $\frac{1}{\sqrt{6}}|\frac{3}{2}, -\frac{1}{2}\rangle+\frac{1}{\sqrt{6}}(1-i)|\frac{3}{2}, \frac{1}{2}\rangle+\frac{1}{\sqrt{2}}i|\frac{3}{2}, \frac{3}{2}\rangle$\\
      &   &   & 2 & $\frac{1}{\sqrt{2}}|\frac{3}{2}, -\frac{3}{2}\rangle+\frac{1}{\sqrt{6}}(1-i)|\frac{3}{2}, -\frac{1}{2}\rangle+\frac{1}{\sqrt{6}}i|\frac{3}{2}, \frac{1}{2}\rangle$\\
$K_1$ & $\frac{3}{2}$ & 2 &  & $-(\frac{\sqrt{3}}{6}(1+i)-\frac{\sqrt{6}}{12}(1-i))|\frac{3}{2}, -\frac{3}{2}\rangle+\frac{1}{2}|\frac{3}{2}, -\frac{1}{2}\rangle-\frac{\sqrt{2}}{4}(1+i)|\frac{3}{2}, \frac{1}{2}\rangle-(\frac{\sqrt{6}}{6}+\frac{\sqrt{3}}{6}i)|\frac{3}{2}, \frac{3}{2}\rangle$\\
$K_2$ & $\frac{3}{2}$ & 2 &  & $-(\frac{\sqrt{3}}{6}(1+i)+\frac{\sqrt{6}}{12}(1-i))|\frac{3}{2}, -\frac{3}{2}\rangle+\frac{1}{2}|\frac{3}{2}, -\frac{1}{2}\rangle+\frac{\sqrt{2}}{4}(1+i)|\frac{3}{2}, \frac{1}{2}\rangle+(\frac{\sqrt{6}}{6}-\frac{\sqrt{3}}{6}i)|\frac{3}{2}, \frac{3}{2}\rangle$\\
\hline
\end{tabular}
\end{center}
\caption{The basis vectors
 of the irreducible representations $\Gamma$ of the little group
 $^2C_{3v}$ and half-integer spin. $\alpha$ labels the components of
 the basis vectors of the two-dimensional representation $G_1$.} 
\label{Table:Basis_Ferm_DiffM_111}
%\end{sidewaystable}
\end{table}

The matrix elements of $\hat{M}$ in the new basis are given by
\begin{equation}
\langle \Gamma\alpha ln |\hat{M} | \Gamma^\prime \alpha^\prime
l^\prime n^\prime \rangle 
= 
\sum_{m m^\prime}c_{lm}^{\,\Gamma \alpha n\,*} \,
c^{\,\Gamma^\prime \alpha^\prime n^\prime}_{l^\prime
  m^\prime}\,\,M_{lm,l^\prime m^\prime} 
\end{equation}
for meson resonances and
\begin{equation}
\langle \Gamma\alpha Jln |\hat{M} | \Gamma^\prime\alpha^\prime
J^\prime l^\prime n^\prime \rangle 
= 
\sum_{\mu \mu^\prime}c_{Jl\mu}^{\,\Gamma \alpha n\,*} \,
c^{\,\Gamma^\prime \alpha^\prime n^\prime}_{J^\prime l^\prime
  \mu^\prime}\,\,M_{Jl\mu,J^\prime l^\prime \mu^\prime}
\end{equation}
for baryon resonances.

According to Schur's lemma, $\hat{M}$ is partially 
diagonalized in the new basis,
\begin{equation}\label{eq:Schur}
\begin{split}
\langle \Gamma\alpha l n |\hat{M}| \Gamma^\prime\alpha^\prime l^\prime
n^\prime\rangle 
&=
          \delta_{\Gamma \Gamma^\prime}  \delta_{\alpha \alpha^\prime} 
M^\Gamma_{ln,l^\prime n^\prime}\,,
\\[2mm]
\langle {\Gamma}\alpha Jl n |\hat{M}| \Gamma^\prime\alpha^\prime
J^\prime l^\prime n^\prime\rangle 
&=
          \delta_{\Gamma \Gamma^\prime}  \delta_{\alpha \alpha^\prime} 
M^\Gamma_{J l n,J^\prime l^\prime n^\prime}\,.
\end{split}
\end{equation}
The phase shift formulae (\ref{boson}) and (\ref{baryon}) then reduce to
\begin{equation}\label{det-eqn1}
\begin{split}
&\det \left( M^\Gamma_{ln,l^\prime n^\prime} -  \delta_{ll^\prime}
  \delta_{n n^\prime}\,\cot\delta_l\,\right)=0\, ,
\\[2mm]
&\det \left( M^\Gamma_{J l n,J^\prime l^\prime n^\prime}-
\delta_{J J^\prime}\delta_{ll^\prime} \delta_{nn^\prime}\, \cot
\delta_{Jl}\right)=0\, . 
\end{split}\end{equation}
The partially diagonalized matrices $M^\Gamma_{ln,l^\prime n^\prime}$ and
$M^\Gamma_{Jln,J^\prime l^\prime n^\prime}$ are given below for $l\leq
2$ and $J=1/2$ and $3/2$.

\subsection{Reduced matrices $\mathbold{M^\Gamma}$}

\subsubsection{$\vd=(0,0,1)$ -- integer spin} 

In this case $N(\Gamma,l)=1$ for all representations, so that we may
drop the subscripts $n, n^\prime$ from $M^\Gamma_{ln,l^\prime
  n^\prime}$. The matrices $M^\Gamma$ have the following entries
\begin{equation}
\left(M^\Gamma_{l,l^\prime}\right) = \left(\begin{tabular}{ccc}
$M_{0,0}$ & $M_{0,1}$ & $M_{0,2}$\\
$M_{1,0}$ & $M_{1,1}$ & $M_{1,2}$\\
$M_{2,0}$ & $M_{2,1}$ & $M_{2,2}$
\end{tabular}\right) 
\end{equation}
with
\begin{eqnarray}
M^{A_1}=
 \begin{pmatrix}
  w_{0 0} & \sqrt{3} i w_{1 0} & -\sqrt{5} w_{2 0} \\
  -\sqrt{3} i w_{1 0} &  w_{0 0}+2 w_{2 0}   & i \frac{2 \sqrt{15}}{5} w_{1 0} + i \frac{3 \sqrt{15}}{5} w_{3 0}  \\
  -\sqrt{5} w_{2 0}  &  -i \frac{2 \sqrt{15}}{5} w_{1 0} - i \frac{3 \sqrt{15}}{5} w_{3 0} &  w_{0 0}+ \frac{10}{7} w_{2 0} + \frac{18}{7} w_{4 0} \\
 \end{pmatrix}\, ,
\end{eqnarray}
\begin{eqnarray}
M^{B_1}=
 \begin{pmatrix}
  0 & 0 & 0 \\
  0 & 0 & 0 \\
  0 & 0 & w_{0 0} - \frac{10}{7} w_{2 0} + \frac{3}{7} w_{4 0} + \frac{3 \sqrt{70}}{7} w_{4 4} \\
 \end{pmatrix}\, ,
\end{eqnarray}
\begin{eqnarray}
{\cal M}^{B_2}=
 \begin{pmatrix}
  0 & 0 & 0 \\
  0 & 0 & 0 \\
  0 & 0 & w_{0 0} - \frac{10}{7} w_{2 0} + \frac{3}{7} w_{4 0} - \frac{3 \sqrt{70}}{7} w_{4 4} \\
 \end{pmatrix}\, ,
\end{eqnarray}
\begin{eqnarray}
M^{E}=
 \begin{pmatrix}
0 & 0 & 0\\
0 & w_{0 0}-w_{2 0} & -\frac{3\sqrt{10}}{10}(1-i)(w_{1 0}-w_{3 0})\\
0 & -\frac{3\sqrt{10}}{10}(1+i)(w_{1 0}-w_{3 0}) & w_{0 0}+\frac{5}{7}w_{2 0}-\frac{12}{7}w_{4 0}
 \end{pmatrix}\, .
\end{eqnarray}

\subsubsection{$\vd=(0,0,1)$ -- half-integer spin} 

In this case $N(\Gamma,l)=1$ for all representations, so that we may
drop the subscripts $n, n^\prime$ again. The matrices
$M^\Gamma$ have the following entries
\begin{equation}
\left(M^\Gamma_{Jl,J^\prime l^\prime}\right) = \left(\begin{tabular}{cccc}
$M_{\frac{1}{2}0,\frac{1}{2}0}$ & $M_{\frac{1}{2}0,\frac{1}{2}1}$ &
  $M_{\frac{1}{2}0,\frac{3}{2}1}$ &$M_{\frac{1}{2}0,\frac{3}{2}2}$  \\
$M_{\frac{1}{2}1,\frac{1}{2}0}$ & $M_{\frac{1}{2}1,\frac{1}{2}1}$ &
  $M_{\frac{1}{2}1,\frac{3}{2}1}$ &$M_{\frac{1}{2}1,\frac{3}{2}2}$  \\
$M_{\frac{3}{2}1,\frac{1}{2}0}$ & $M_{\frac{3}{2}1,\frac{1}{2}1}$ &
  $M_{\frac{3}{2}1,\frac{3}{2}1}$ &$M_{\frac{3}{2}1,\frac{3}{2}2}$  \\
$M_{\frac{3}{2}2,\frac{1}{2}0}$ & $M_{\frac{3}{2}2,\frac{1}{2}1}$ &
  $M_{\frac{3}{2}2,\frac{3}{2}1}$ &$M_{\frac{3}{2}2,\frac{3}{2}2}$ 
\end{tabular}\right) 
\end{equation}
with
\begin{eqnarray}
M^{G_1}=
 \begin{pmatrix}
w_{0 0} & i w_{1 0} & i\sqrt{2} w_{1 0} & -\sqrt{2} w_{2 0}\\
-i w_{1 0} & w_{0 0} & \sqrt{2} w_{2 0} & i\sqrt{2} w_{1 0}\\
-i\sqrt{2} w_{1 0} & \sqrt{2} w_{2 0} & w_{0 0}+w_{2 0} & \frac{i}{5}w_{1 0}+\frac{9i}{5}w_{3 0}\\
-\sqrt{2} w_{2 0} & -i\sqrt{2} w_{1 0} & -\frac{i}{5}w_{1 0}-\frac{9i}{5}w_{3 0} & w_{0 0}+w_{2 0}
\end{pmatrix}\, ,
\end{eqnarray}
\begin{eqnarray}
M^{G_2}=
 \begin{pmatrix}
0 & 0 & 0 & 0\\
0 & 0 & 0 & 0\\
0 & 0 & w_{0 0}-w_{2 0} & i\frac{3}{5}(w_{1 0}-w_{3 0})\\
0 & 0 & -i\frac{3}{5}(w_{1 0}-w_{3 0}) & w_{0 0}-w_{2 0}
\end{pmatrix}\, .
\end{eqnarray}

\subsubsection{$\vd=(1,1,0)$ -- integer spin} 

The representation $A_1$ occurs twice in $D^l$ for $l=2$, $N(A_1,2) =
2$. In all other cases $N(\Gamma,l) = 1$. The matrices
$M^\Gamma$ have the following entries
\begin{equation}
\left(M^\Gamma_{l[n],l^\prime[n^\prime]}\right) = \left(\begin{tabular}{cccc}
$M_{0,0}$ & $M_{0,1}$ & $M_{0,21}$ & $M_{0,22}$\\
$M_{1,0}$ & $M_{1,1}$ & $M_{1,21}$ & $M_{1,22}$\\
$M_{21,0}$ & $M_{21,1}$ & $M_{21,21}$ & $M_{21,22}$ \\
$M_{22,0}$ & $M_{22,1}$ & $M_{22,21}$ & $M_{22,22}$ 
\end{tabular}\right) 
\end{equation}
with
\begin{eqnarray}
{\cal M}^{A_1}=
 \begin{pmatrix}
w_{0 0} & -\sqrt{6}(1-i) r_{11} & \sqrt{10} w_{2 2} & -\sqrt{5} w_{2 0}\\
-\sqrt{6}(1+i) r_{11} & A_{22} & A_{23}^- & A_{24}^+\\
-\sqrt{10} w_{2 2} & A_{23}^+ & A_{33} & A_{34}\\
-\sqrt{5} w_{2 0} & A_{24}^-& -A_{34} & A_{44}
\end{pmatrix}\, ,
\end{eqnarray}
where $r_{js}=\mbox{Re}\,w_{js}$ and 
\begin{equation}
\begin{split}
A_{23}^\pm&=\frac{3 \sqrt{10}}{5} (1\pm i) r_{1 1}-\frac{\sqrt{15}}{5}(1\pm i)r_{3 1}-3(1\pm i)r_{3 3}\, ,
\\[2mm]
A_{24}^\pm&=-\frac{\sqrt{30}}{5}(1\pm i)r_{1 1}+\frac{6 \sqrt{5}}{5}(1\pm i)r_{3 1}\, ,
\\[2mm]
A_{22}&=w_{0 0}-w_{2 0}-\sqrt{6} i w_{2 2} \,,
\\[2mm]
A_{33}&=
w_{0 0}-\frac{10}{7}w_{2 0}+ \frac{3}{7}w_{4 0}- \frac{3 \sqrt{70}}{7}w_{4 4}\, ,
\\[2mm]
A_{34}&=-\frac{10 \sqrt{2}}{7} w_{2 2}+\frac{3 \sqrt{30}}{7} w_{42}\, , 
\\[2mm]
A_{44}&=w_{0 0}+\frac{10}{7}w_{2 0}+ \frac{18}{7}w_{4 0} \,.
\end{split}
\end{equation}
For the remaining representations we drop the subscripts $n, n^\prime$
and obtain
\begin{eqnarray}
M^{A_2}=
 \begin{pmatrix}
0 & 0 & 0\\
0 & 0 & 0\\
0 & 0 & w_{0 0}+\frac{5}{7} w_{2 0}-\frac{12}{7} w_{4 0}+\frac{5 \sqrt{6}}{7} i w_{2 2}+\frac{6 \sqrt{10}}{7} i w_{4 2}
\end{pmatrix}\, ,
\end{eqnarray}
\begin{eqnarray}
M^{B_1}=
 \begin{pmatrix}
0 & 0 & 0\\
0 & w_{0 0} + 2 w_{2 0} & -\frac{3\sqrt{10}}{5}(1+i)r_{1 1}-\frac{4 \sqrt{15}}{5}(1+i)r_{3 1}\\
0 & -\frac{3\sqrt{10}}{5}(1-i)r_{1 1}-\frac{4 \sqrt{15}}{5}(1-i)r_{3 1} & B_{33}
\end{pmatrix}\, ,
\end{eqnarray}
\begin{eqnarray}
M^{B_2}=
 \begin{pmatrix}
0 & 0 & 0\\
0 & w_{0 0} - w_{2 0} + \sqrt{6}i w_{2 2} & \tilde B_{23}^-\\
0 & \tilde B_{23}^+& w_{0 0}-\frac{10}{7} w_{2 0}+\frac{3}{7} w_{4 0}+ \frac{3 \sqrt{70}}{4} w_{4 4}
\end{pmatrix}\, ,
\end{eqnarray}
where
\begin{equation}
\begin{split}
B_{33}&=w_{0 0}+\frac{5}{7} w_{2 0}-\frac{12}{7} w_{4 0}-\frac{5
  \sqrt{6}}{7} i w_{2 2}-\frac{6 \sqrt{10}}{7} i w_{4 2} \, ,\\[2mm]
\tilde B_{23}^\pm&=
-\frac{3\sqrt{10}}{5}(1\pm i)r_{1 1}+\frac{\sqrt{15}}{5}(1\pm i)r_{3 1}-3(1\pm i)r_{3 3}\, .
\end{split}
\end{equation}

\subsubsection{$\vd=(1,1,0)$ -- half-integer spin} 

In this case we are concerned with the representation $G_1$ only. We
have $N(G_1,l)=2$ for $J=\frac{3}{2}$ and $l=1, 2$ and $N(G_1,l)=1$
else. The matrix $M^{G_1}$ has the following entries
\begin{equation}
\left(M^{G_1}_{Jl[n],J^\prime l^\prime[n^\prime]}\right) = \left(\begin{tabular}{cccccc}
$M_{\frac{1}{2}0,\frac{1}{2}0}$ & $M_{\frac {1}{2}0,\frac{1}{2}1}$ &
  $M_{\frac{1}{2}0,\frac{3}{2}11}$ &$M_{\frac{1}{2}0,\frac{3}{2}12}$ 
  $M_{\frac{1}{2}0,\frac{3}{2}21}$ &$M_{\frac{1}{2}0,\frac{3}{2}22}$  \\
$M_{\frac{1}{2}1,\frac{1}{2}0}$ & $M_{\frac{1}{2}1,\frac{1}{2}1}$ &
  $M_{\frac{1}{2}1,\frac{3}{2}11}$ &$M_{\frac{1}{2}1,\frac{3}{2}12}$ 
  $M_{\frac{1}{2}1,\frac{3}{2}21}$ &$M_{\frac{1}{2}1,\frac{3}{2}22}$  \\
$M_{\frac{3}{2}11,\frac{1}{2}0}$ & $M_{\frac{3}{2}11,\frac{1}{2}1}$ &
  $M_{\frac{3}{2}11,\frac{3}{2}11}$ &$M_{\frac{3}{2}11,\frac{3}{2}12}$ 
  $M_{\frac{3}{2}11,\frac{3}{2}21}$ &$M_{\frac{3}{2}11,\frac{3}{2}22}$  \\
$M_{\frac{3}{2}12,\frac{1}{2}0}$ & $M_{\frac{3}{2}12,\frac{1}{2}1}$ &
  $M_{\frac{3}{2}12,\frac{3}{2}11}$ &$M_{\frac{3}{2}12,\frac{3}{2}12}$ 
  $M_{\frac{3}{2}12,\frac{3}{2}21}$ &$M_{\frac{3}{2}12,\frac{3}{2}22}$  \\
$M_{\frac{3}{2}21,\frac{1}{2}0}$ & $M_{\frac{3}{2}21,\frac{1}{2}1}$ &
  $M_{\frac{3}{2}21,\frac{3}{2}11}$ &$M_{\frac{3}{2}21,\frac{3}{2}12}$ 
  $M_{\frac{3}{2}21,\frac{3}{2}21}$ &$M_{\frac{3}{2}21,\frac{3}{2}22}$  \\
$M_{\frac{3}{2}22,\frac{1}{2}0}$ & $M_{\frac{3}{2}22,\frac{1}{2}1}$ &
  $M_{\frac{3}{2}22,\frac{3}{2}11}$ &$M_{\frac{3}{2}22,\frac{3}{2}12}$ 
  $M_{\frac{3}{2}22,\frac{3}{2}21}$ &$M_{\frac{3}{2}22,\frac{3}{2}22}$
\end{tabular}\right) 
\end{equation}
with
\begin{eqnarray}
M^{G_1}=
 \begin{pmatrix}
w_{0 0} & -\sqrt{2}G_{14}^- & -\sqrt{3}G_{14}^+ & G_{14}^- & -2 i w_{2 2} & -\sqrt{2} w_{2 0}\\
-\sqrt{2}G_{14}^+ & w_{0 0} & 2 w_{2 2} & \sqrt{2} w_{2 0} & -\sqrt{3}G_{14}^+ & -G_{14}^+\\
-\sqrt{3}G_{14}^- & -2 w_{2 2} & w_{0 0}-w_{2 0} & \sqrt{2} w_{2 2} & G_{35}^- & G_{36}^-\\
G_{14}^+ & \sqrt{2} w_{2 0} & -\sqrt{2} w_{2 2} & w_{0 0}+w_{2 0} & -G_{36}^+ & G_{46}^+\\
-2 i w_{2 2} & -\sqrt{3}G_{14}^- & G_{35}^+ & -G_{36}^- & w_{0 0}-w_{2 0} & -\sqrt{2} i w_{2 2}\\
-\sqrt{2} w_{2 0} & -G_{14}^- & G_{36}^+ & G_{46}^- & -\sqrt{2} i w_{2 2} & w_{0 0}+w_{2 0}
\end{pmatrix}\, ,
\end{eqnarray}
where 
\begin{equation}
\begin{split}
G_{14}^\pm&=(1 \pm i)r_{1 1}\, ,
\\[2mm]
G_{35}^\pm&=\frac{6 \sqrt{5}}{5}(1\pm i)r_{3 3}\, ,
\\[2mm]
G_{36}^\pm&=
-\frac{\sqrt{6}}{5}(1\pm i)(r_{1 1}-\sqrt{6}r_{3 1})\, ,
\\[2mm]
G_{46}^\pm&=-\frac{2 \sqrt{2}}{5}(1\pm i)r_{1 1}-\frac{6
  \sqrt{3}}{5}(1\pm i)r_{3 1} \,.
\end{split}
\end{equation}

\subsubsection{$\vd=(1,1,1)$ -- integer spin} 

In this case we are concerned with two representations, $A_1$ and
$E$. The representation $A_1$ occurs only once in $D^l$, and we find
\begin{equation}
\begin{split}
M^{A_1} &= \left(M^{A_1}_{l,l^\prime}\right) = \left(\begin{tabular}{ccc}
$M_{0,0}$ & $M_{0,1}$ & $M_{0,2}$\\
$M_{1,0}$ & $M_{1,1}$ & $M_{1,2}$\\
$M_{2,0}$ & $M_{2,1}$ & $M_{2,2}$
\end{tabular}\right) \\[2mm]
&=
 \begin{pmatrix}
  w_{0 0} & \frac{3\sqrt{2}}{2} (1-i) w_{1 0}& \sqrt{30} w_{2 2} \\
  \frac{3\sqrt{2}}{2} (1+i) w_{1 0} & w_{0 0} - 2 \sqrt{6} i w_{2 2} & 
\tilde A_{23}^-\\
  - \sqrt{30} w_{2 2} & \tilde A_{23}^+ &  w_{0 0} - \frac{12}{7} w_{4 0}
- \frac{12 \sqrt{10}}{7} i w_{4 2} - \frac{10 \sqrt{6}}{7} i w_{2 2}
 \\
 \end{pmatrix}\, ,
\end{split}
\end{equation}
where
\eq
\tilde A_{23}^\pm=- \frac{3 \sqrt{10}}{5} (1\pm i) w_{1 0} 
+ \frac{3 \sqrt{10}}{5} (1\pm i) w_{3 0} \mp \sqrt{3} (1\mp i) w_{32}\, .
\en
The representation $E$ occurs twice in $D^l$ for $l=2$, $N(E,2)=2$,
and we obtain
\begin{equation}
\begin{split}
M^{E} &= \left(M^{E}_{l[n],l^\prime[n^\prime]}\right) =
\left(\begin{tabular}{cccc} 
$M_{0,0}$ & $M_{0,1}$ & $M_{0,21}$ & $M_{0,22}$\\
$M_{1,0}$ & $M_{1,1}$ & $M_{1,21}$ & $M_{1,22}$\\
$M_{21,0}$ & $M_{21,1}$ & $M_{21,21}$ & $M_{21,22}$\\
$M_{22,0}$ & $M_{22,1}$ & $M_{22,21}$ & $M_{22,22}$
\end{tabular}\right) \\[2mm]
&=
 \begin{pmatrix}
 0  &  0  &  0  &  0  \\
 0  &  w_{0 0} + i \sqrt{6} w_{2 2} &E_{23}^- & E_{24}\\
 0  &  E_{23}^+ &  w_{0 0} + \frac{18}{7} w_{4 0} & E_{34}^-\\
 0  &  E_{24} & -E_{34}^+ & w_{0 0} - \frac{12}{7} w_{4 0} +\frac{6 \sqrt{10}}{7} i w_{4 2} + \frac{5 \sqrt{6}}{7} i w_{2 2}  \\
\end{pmatrix}\, ,
\end{split}
\end{equation}
where
\begin{equation}
\begin{split}
E_{23}^\pm&=\frac{3 \sqrt{5}}{5} (1\pm i) w_{1 0} + \frac{9 \sqrt{5}}{10} (1\pm i) w_{3 0}\, ,
\\[2mm] 
E_{24}&=\frac{3 \sqrt{5}}{5} (w_{1 0}-w_{3 0}) + \sqrt{6} i w_{32} \, ,
\\[2mm] 
E_{34}^\pm&=
\frac{5 \sqrt{6}}{7} (1\pm i) w_{2 2} - \frac{9 \sqrt{10}}{14} (1\pm
i) w_{4 2} \, .
\end{split}
\end{equation}

\subsubsection{$\vd=(1,1,1)$ -- half-integer spin} 

In this case $N(\Gamma,l)=1$ for all representations. The 
matrices $M^\Gamma$ have the following entries
\begin{equation}
\left(M^\Gamma_{Jl,J^\prime l^\prime}\right) = \left(\begin{tabular}{cccc}
$M_{\frac{1}{2}0,\frac{1}{2}0}$ & $M_{\frac{1}{2}0,\frac{1}{2}1}$ &
  $M_{\frac{1}{2}0,\frac{3}{2}1}$ &$M_{\frac{1}{2}0,\frac{3}{2}2}$  \\
$M_{\frac{1}{2}1,\frac{1}{2}0}$ & $M_{\frac{1}{2}1,\frac{1}{2}1}$ &
  $M_{\frac{1}{2}1,\frac{3}{2}1}$ &$M_{\frac{1}{2}1,\frac{3}{2}2}$  \\
$M_{\frac{3}{2}1,\frac{1}{2}0}$ & $M_{\frac{3}{2}1,\frac{1}{2}1}$ &
  $M_{\frac{3}{2}1,\frac{3}{2}1}$ &$M_{\frac{3}{2}1,\frac{3}{2}2}$  \\
$M_{\frac{3}{2}2,\frac{1}{2}0}$ & $M_{\frac{3}{2}2,\frac{1}{2}1}$ &
  $M_{\frac{3}{2}2,\frac{3}{2}1}$ &$M_{\frac{3}{2}2,\frac{3}{2}2}$ 
\end{tabular}\right) 
\end{equation}
with
\begin{eqnarray}
M^{K_1}=
 \begin{pmatrix}
0 & 0 & 0 & 0\\
0 & 0 & 0 & 0\\
0 & 0 & w_{0 0}+\sqrt{6}i w_{2 2} & \bar B_{34}^+\\
0 & 0 & \bar B_{34}^- & w_{0 0}+\sqrt{6}i w_{2 2}
\end{pmatrix}\, ,
\end{eqnarray}
where
\eq
\bar B_{34}^\pm=
\left(-\frac{3\sqrt{2}}{5}\pm \frac{3}{5}i\right)w_{1 0}+\left(\frac{3\sqrt{2}}{5}\pm\frac{12}{5}i\right)w_{3 0}+\left(\pm\frac{\sqrt{30}}{5}-\frac{2\sqrt{15}}{5}i\right)w_{3 2}\, ,
\en
and
\begin{eqnarray}
M^{K_2}=
 \begin{pmatrix}
0 & 0 & 0 & 0\\
0 & 0 & 0 & 0\\
0 & 0 & w_{0 0}+\sqrt{6}i w_{2 2} & \hat B_{34}^+\\
0 & 0 & \hat B_{34}^- & w_{0 0}+\sqrt{6}i w_{2 2}
\end{pmatrix}\, ,
\end{eqnarray}
where
\eq
\hat B_{34}^\pm= \left(\frac{3\sqrt{2}}{5}\pm \frac{3}{5}i\right)w_{1 0}
+\left(-\frac{3\sqrt{2}}{5}\pm\frac{12}{5}i\right)w_{3 0}
+\left(\pm \frac{\sqrt{30}}{5}+\frac{2\sqrt{15}}{5}i\right)w_{3 2}\, ,
\en
and finally
\begin{eqnarray}
M^{G_1}=
 \begin{pmatrix}
w_{0 0} & \frac{\sqrt{6}}{2}(1-i)w_{1 0} & \sqrt{6}i w_{1 0} & -2\sqrt{3}i w_{2 2}\\
\frac{\sqrt{6}}{2}(1+i)w_{1 0} & w_{0 0} & \sqrt{6}(1+i)w_{2 2} & \sqrt{3}(1+i)w_{1 0}\\
-\sqrt{6}i w_{1 0} & -\sqrt{6}(1-i)w_{2 2} & w_{0 0}-i\sqrt{6} w_{2 2} & 
\tilde G_{34}\\
-2\sqrt{3}i w_{2 2} & \sqrt{3}(1-i)w_{1 0} & -\tilde G_{34}
 & w_{0 0}-i\sqrt{6} w_{2 2}
\end{pmatrix}\, ,
\end{eqnarray}
where
\eq
\tilde G_{34}=\frac{6\sqrt{3}}{5}i w_{3 0}-\frac{\sqrt{3}}{5}i w_{1 0}-\frac{3\sqrt{10}}{5} w_{3 2}\, .
\en

\section{Three examples}
\label{secex}

\begin{table}[b]
\begin{center}
\begin{tabular}{|c|c|c|c|c|}
\hline
$\vd$ & Little Group & $\Gamma$ & $\cot \delta_1$ \\
\hline
\multirow{2}{*}{$(0,0,1)$} & \multirow{2}{*}{$C_{4v}$}& $A_1^\pm$ &
$w_{00}+2 w_{20}$\\
 & & $E^\pm$ &
$w_{00}- w_{20}$\\ \hline
\multirow{3}{*}{$(1,1,0)$} & \multirow{3}{*}{$C_{2v}$}& $A_1^\pm$ &
$w_{00}- w_{20}-i \sqrt{6} w_{22}$\\
& & $B_1^\pm$ &
$w_{00}+ 2 w_{20}$\\
& & $B_2^\pm$ &
$w_{00}- w_{20}+i \sqrt{6} w_{22}$\\ \hline
\multirow{2}{*}{$(1,1,1)$} & \multirow{2}{*}{$C_{3v}$} & $A_1^\pm$&
$w_{00}- i 2\sqrt{6} w_{22}$\\
& & $E^\pm$&
$w_{00}+ i \sqrt{6} w_{22}$\\
\hline
\end{tabular}
\end{center}
\caption{The phase shifts of the $\rho$ resonance
  for the various boost vectors and representations.}
\label{Table:rho}
\end{table}

Let us now apply the formulae derived above to a few
concrete cases. A general feature of unequal mass particles is that
spin and angular momentum mix under the Lorentz boost, which
complicates the extraction of phase shifts significantly. In the case
of baryon resonances, nonvanishing momenta prove most advantageous for
the evaluation of $P$-wave phase shifts, as we have seen in the
Introduction. For $S$-wave baryon resonances nonvanishing momenta are
of no big advantage as far as moving the level crossing to smaller
values of $m_\pi L$ is concerned.

Of primary interest are the $\rho$ and the $\Delta$ resonance. The
calculation of the mass and the width  
of the $\rho$ meson provides a benchmark test, which has to be passed
successfully before we can address more complex systems. The $\Delta$
resonance is interesting for two reasons. First of all, it is one of
the very few elastic two-body baryon resonances, and as such qualifies
for a first extension of L\"uscher's method to particles carrying
spin. Secondly, being a $P_{33}$ wave, its phase $\delta_{\frac{3}{2} 1}$ can
be computed directly from representations $G_2$ and $B_1$,
$B_2$. Finally, we consider the $N^\star(1440)$ Roper resonance. Being a
$P_{11}$ wave and carrying spin $1/2$, it couples to the
representation $G_1$ only, which mixes 
spin $1/2$ with spin $3/2$ and angular momentum $l=0$ with angular
momentum $l=1$.

\subsection{The $\mathbold{\rho}$ resonance}

In the case of equal masses and integer spin the situation
simplifies significantly. All matrices $M^\Gamma$ turn out
to be diagonal, and the phase shifts can be directly read off from
their eigenvalues. The phase shifts $\delta_1$ of the $\rho$ resonance
are given in Table~\ref{Table:rho}. 

\subsection{The $\mathbold{\Delta(1232)}$ resonance}

Neglecting mixing with D waves (and higher), it is straightforward to
compute $\delta_{\frac{3}{2} 1}$ for boost vector $\vd=(0,0,1)$ from
the representation $G_1$ and for boost vector
$\vd=(1,1,1)$ from representations $B_1$ and $B_2$, giving
\begin{equation}
\begin{tabular}{|c|c|c|} \hline
$\vd$ & $\Gamma$ & $\cot \delta_{\frac{3}{2} 1}$ \\ \hline
$(0,0,1)$ & $G_2$ & $w_{00} - w_{20}$ \\ \hline
$(1,1,1)$ & $B_1, B_2$ & $w_{00} + i \sqrt{6} w_{22}$ \\ \hline
\end{tabular}
\end{equation}
In all other cases $\delta_{\frac{3}{2} 1}$ mixes with lower spin and
lower partial waves.

The same formulae apply to the $\Sigma^*(1385)$ resonance (whose
energy levels we have shown in Fig.~\ref{figS}), which is a 
$P_{13}$ wave. 

\subsection{The $\mathbold{N^\star(1440)}$ Roper resonance}

Let us consider the boost vector $\vd=(0,0,1)$ and representation
$G_1$. Alternatively we could 
consider the boost vectors $\vd=(1,1,0)$ and $\vd=(1,1,1)$. Neglecting
mixing with $J=\frac{3}{2}$ states for the moment, we need to solve
\begin{equation}
\left|\begin{tabular}{cc}
$w_{00}-\cot \delta_{\frac{1}{2} 0}$ & $i\,w_{10}$ \\
$-i\,w_{10}$ & $w_{00}-\cot \delta_{\frac{1}{2} 1}$
\end{tabular}\right| = 0 \,,
\label{roper}
\end{equation}
which leads to
\begin{equation}
\cot \delta_{\frac{1}{2} 1}\,\cot \delta_{\frac{1}{2} 0} 
- \left(\cot \delta_{\frac{1}{2} 1} -\cot \delta_{\frac{1}{2}
  0}\right)\,w_{00} 
+w_{00}^2 - w_{10}^2 = 0 \,.
\label{roper2}
\end{equation}
The phase shift that interests us here is $\delta_{\frac{1}{2} 1}$. To
compute $\delta_{\frac{1}{2} 1}$ from (\ref{roper2}) we need to know
$\delta_{\frac{1}{2} 0}$, which is most easily obtained from
eigenstates of zero total momentum, $\vd=(0,0,0)$.
It is not excluded that the spin-$1/2$ states mix with the $P$-wave
spin-$3/2$ state, though no resonance of that kind has been reported
by the Particle Data Group~\cite{pdg}. In this case we would have
\begin{equation}
\left|\begin{tabular}{ccc}
$w_{00}-\cot \delta_{\frac{1}{2} 0}$ & $i\,w_{10}$ & $i\,\sqrt{2}\,w_{10}$\\
$-i\,w_{10}$ & $w_{00}-\cot \delta_{\frac{1}{2} 1}$ &
$\sqrt{2}\,w_{20}$ \\
$-i\,\sqrt{2}\,w_{10}$ & $\sqrt{2}\,w_{20}$ & $w_{00} + w_{20} - \cot
\delta_{\frac{3}{2} 1}$ 
\end{tabular}\right| = 0 \,.
\label{beyondroper}
\end{equation}
To find out, and to solve (\ref{beyondroper}), the phase
$\delta_{\frac{3}{2} 1}$ can be directly computed from representation
$G_2$. It has to be extrapolated to the appropriate value of $q^2$
though. 

\section{Operators}
\label{sec6}

Recently, several
authors~\cite{Thomas:2011rh,Leskovec:2012gb,Dudek:2012gj,Foley:2012wb}
have started to construct operators projecting onto selected irreducible
representations of the little groups. In this
Section we extend the work to higher representations and/or particles with
spin.    

We start from (generally nonlocal) operators
$O_\alpha(\vx_1,\vx_2,t)$. Under space rotations $\hat{R}$ they transform like
\eq
(\hat{R}\,O)_\alpha(\vx_1,\vx_2,t)
=S_{\alpha\beta}(R)\,O_\beta(R^{-1}\vx_1,R^{-1}\vx_2,t)\, ,
\en
where $R^{-1}\vx$ denotes the rotated vector $\vx$, and 
the matrices $S_{\alpha\beta}(R)$ form a linear representation of the
group $SO(3)$ in case of integer spin and $SU(2)$ in case of
half-integer spin. The explicit form of $S_{\alpha\beta}(R)$ is
well known for scalar, vector and spinor fields.
Under space inversions $\hat{I}$ the operators transform as
\eq
(\hat{I}\,O)_\alpha(\vx_1,\vx_2,t)
=I_{\alpha\beta}\,O_\beta(-\vx_1,-\vx_2,t)
\en
with $I^2=1$. 
An operator $O^\Gamma_\alpha(\vx_1,\vx_2,t)$, which transforms
according to the irreducible representation $\Gamma$ of the little
group, is given by 
(see, for example,~\cite{Elliott})
\eq\label{pro}
O^\Gamma_\alpha(\vx_1,\vx_2,t)
=\sum_i\chi_\Gamma^*(S_i)\,(\hat{S}_i\,O)_\alpha(\vx_1,\vx_2,t)\, ,
\en
where the sum runs over all elements $S_i$ of the little group, which
are either pure rotations $R_i$, or rotations combined with space inversion
$I\,R_i$. The quantities $\chi_\Gamma(S_i)$ denote the characters in the
representation $\Gamma$. 
The operators can be trivially Fourier transformed to momentum
space. 

Below we will give a few examples of single-particle and two-particle
operators, which demonstrate the procedure to be followed in the
general case.

\subsection{Single-particle operators}

Let us start with the simple case of quark-antiquark and three-quark
operators, and discuss this case in detail.

\subsubsection{The case $\vd=(0,0,1)$ -- scalar mesons}

Consider the operator
\eq
O(\vp,t)=\sum_{\vx}e^{i \vp \vx}\,\bar{q}(\vx,t)q(\vx,t)\, .
\en
Under rotations and space inversions the operator transforms as 
\eq
\hat{R}\left(\bar{q}(\vx,t)q(\vx,t)\right)=\bar{q}(R^{-1}\vx,t)q(R^{-1}\vx,t)\, ,\quad
\hat{I}\left(\bar{q}(\vx,t)q(\vx,t)\right)=\bar{q}(-\vx,t)q(-\vx,t)\, .
\en
The projected operator takes the form
\eq
O^\Gamma(\vp,t)=\sum_i\chi_\Gamma^*(S_i)
\sum_{\vx}e^{i \vp \vx}\,\bar{q}(S^{-1}_i\vx,t)q(S^{-1}_i\vx,t)\, .
\en
Note that the sites $R_i^{-1}\vx$ and $-\vx$ belong to the lattice if $\vx$
does. In the case of unequal masses, 
the momentum $\vp$ is left invariant by the elements of the little group,
$S_i\,\vp=\vp$, so that we have
\eq\label{eq:scalar-unequal}
O^\Gamma(\vp,t)=\sum_i\chi_\Gamma^*(S_i)
\sum_{\vx}e^{i \vp \vx}\,\bar{q}(\vx,t)q(\vx,t)
=\left(\sum_i\chi_\Gamma^*(S_i)\right)\,O(\vp,t)\, .
\en
Consequently, the operator $O(\vp,t)$ transforms according to the
trivial representation 
$A_1$, for which $\sum_i\chi_\Gamma^*(S_i)\neq 0$.

In the case of equal masses, the number of irreducible
representations is doubled, $\Gamma \rightarrow \Gamma^\pm$, and the
momentum $\vp$ is left invariant by the elements of the little group
up to a sign, $S_i\, \vp=\pm \vp$. Accordingly, the operators
$O^\Gamma$ should be symmetrized, or antisymmetrized, with respect to
$\vp \leftrightarrow -\vp$.
%Thus, (\ref{eq:scalar-unequal}) is to be replaced by  
%\eq
%O^{\Gamma^\pm}(\vp,t)=\left(\sum_i\chi_\Gamma^*(S_i)\right)
%\sum_{\vx}(e^{i \vp \vx}\pm e^{-i \vp \vx})\,\bar{q}(\vx,t)q(\vx,t)\, ,
%\en
%where the sum runs over those $S_i$, which
%obey $S_i^{-1}\vp=\vp$, as in (\ref{eq:scalar-unequal}).
However, we may still work with the same
operators as for unequal masses. The advantage of these operators is
that they have 
definite momentum $\vp$. The additional symmetry present in
the case of equal masses has solely the effect that even angular
momenta do not mix with odd angular 
momenta in the spectrum of the lattice Hamiltonian.

\subsubsection{The case $\vd=(0,0,1)$ -- vector mesons}

Starting from the operator
\eq\label{eq:vector-ini}
V(\vp,t)= \begin{pmatrix}V_1(\vp,t)\cr V_2(\vp,t)\cr
  V_3(\vp,t)\end{pmatrix} \,\quad V_i(\vp,t)=\sum_{\vx}e^{i \vp
  \vx}\,\bar{q}(\vx,t)\,\gamma_i \,q(\vx,t)\, , 
\en
it can easily be checked that in the case of unequal masses the
operator
\eq\label{eq:E001}
V^E(\vp,t)=
\sum_{\vx}e^{i \vp \vx}
\begin{pmatrix}V_1(\vx,t)\cr V_2(\vx,t)\cr 0\end{pmatrix}
\en
transforms according to the irreducible representation $E$. In fact,
one may use 
any linear combination  of $V_1$ and $V_2$ to project onto $E$. In
contrast, the third component, 
$V_3$, transforms according to the irreducible representation $A_1$, 
\eq\label{eq:A1}
V^{A_1}(\vp,t)=\sum_{\vx}e^{i \vp \vx}\,V_3(\vx,t)\, .
\en
%To arrive at operators $V^{E^\pm}$ and $V^{A_1^\pm}$ for equal masses, we
%have to replace the exponential $e^{i \vp \vx}$ in (\ref{eq:E001}) and
%(\ref{eq:A1}) by $e^{i \vp \vx}\pm e^{-i \vp \vx}$.

\subsubsection{The case $\vd=(1,1,1)$ -- vector mesons}

We start again from the operator (\ref{eq:vector-ini}). Instead of
(\ref{eq:E001}), we now get 
\eq\label{eq:E111}
V^E(\vp,t)=
\sum_{\vx}e^{i \vp \vx}
\begin{pmatrix}2V_1(\vx,t)-V_2(\vx,t)-V_3(\vx,t)\cr
2V_2(\vx,t)-V_3(\vx,t)-V_1(\vx,t)\cr
2V_3(\vx,t)-V_1(\vx,t)-V_2(\vx,t)\end{pmatrix}\, .
\en  
The components of $V^E$ are cyclic permutations, $\{i,j,k\} = \{1,2,3\}$, of
the operator $2V_i-V_j-V_k$. Note that only two of the components of
$V^E$ are independent, in accord with the representation $E$ being
two-dimensional. 

The operator projected onto the prepresentation $A_1$ is given by
\eq
V^{A_1}(\vp,t)=\sum_{\vx}e^{i \vp \vx}\,(V_1(\vx,t)
+V_2(\vx,t)+V_3(\vx,t))\, .
\en
%In the case of equal masses we again have to replace $e^{i \vp \vx}$
%by $e^{i \vp \vx}\pm e^{-i \vp \vx}$. 

\subsubsection{The case $\vd=(0,0,1)$ -- $\Delta$ resonance} 

We start from the interpolating operator of the $\Delta^+$ resonance,
\begin{equation}
\Delta^+_{\alpha}(\vp,t)=\begin{pmatrix}\Delta^+_{1\alpha}(\vp,t) \cr
\Delta^+_{2\alpha}(\vp,t) \cr \Delta^+_{3\alpha}(\vp,t)\end{pmatrix} 
\end{equation}
with, for example,
\begin{equation}
\begin{split}
\Delta^+_{i\alpha}(\vp,t) &= \sum_{\vx}e^{i \vp \vx}\,\left\{ 
2(u^T(\vx,t)C\gamma_id(\vx,t)u_\alpha(\vx,t)
+(u^T(\vx,t)C\gamma_iu(\vx,t)d_\alpha(\vx,t)\right\}
\\[2mm]
&= \sum_{\vx}e^{i \vp \vx}\Delta^+_{i\alpha}(\vx,t)\, .
\end{split}
\end{equation}
Under space rotations the operator $\Delta^+_{i\alpha}(\vp,t)$ transforms as
\eq
(\hat{R}\Delta)^+_{i\alpha}(\vx,t)
=S_{\alpha\beta}(R)\,A_{ij}(R)\,\Delta^+_{j\beta}(R^{-1}\vx,t)\, ,\quad
S(R)=\begin{pmatrix} \check{S}(R) & 0 \cr 0 & \check{S}(R)\end{pmatrix}\, ,
\en
where $A(R)$ and $\check{S}(R)$ are $3\times 3$ and $2\times 2$
irreducible matrix representations of $SU(2)$, respectively.
Under space inversions the operator transforms as
\eq
(\hat{I}\Delta)^+_{i\alpha}(\vx,t)
=(\gamma_0)_{\alpha\beta}\Delta^+_{i\beta}(-\vx,t)\, .
\en
Applying (\ref{pro}), the operators projected onto the irreducible
representations $G_1$ and $G_2$ turn out to be
\begin{equation}
\Delta^{+G_1}_{\alpha}=\sum_{\vx}e^{i \vp \vx}\,
\begin{pmatrix}
\Delta^+_{1\alpha}(\vx,t)+i(\Sigma_3)_{\alpha\beta}\Delta^+_{2\beta}(\vx,t)\cr
\Delta^+_{2\alpha}(\vx,t)-i(\Sigma_3)_{\alpha\beta}\Delta^+_{1\beta}(\vx,t)\cr
 2\Delta^+_{3\alpha}(\vx,t) 
\end{pmatrix}
\label{twocomp}
\end{equation}
and
\begin{equation}
 \Delta^{+G_2}_{\alpha}=\sum_{\vx}e^{i \vp \vx}\,
\begin{pmatrix}
\Delta^+_{1\alpha}(\vx,t)-i(\Sigma_3)_{\alpha\beta}\Delta^+_{2\beta}(\vx,t)\cr
\Delta^+_{2\alpha}(\vx,t)+i(\Sigma_3)_{\alpha\beta}\Delta^+_{1\beta}(\vx,t)\cr
0
\end{pmatrix}\, ,
\end{equation}
respectively, where
\eq
\Sigma_3=\begin{pmatrix}\sigma_3 & 0 \cr 0 & \sigma_3\end{pmatrix}\, .
\en
%Note again that only two of the components of (\ref{twocomp}) are independent.

\subsection{Two-particle operators}

\subsubsection{The case $\vd=(0,0,1)$ -- product of two (pseudo-)scalar fields}

We start from the operator
\eq
O(\vp,\vq,t)=\sum_{\vx,\vy}e^{i (\vp \vx + \vq (\vx - \vy))}
\phi_1(\vx,t)\,\phi_2(\vy,t)\, .
\en
In the case of unequal masses the operator that transforms according to the
irreducible representation $\Gamma$ is given by
\eq
O^\Gamma(\vp,\vq,t)=\sum_{i=1}^8\chi_\Gamma^*(S_i)
\sum_{\vx, \vy}e^{i (\vp \vx + (S_i\vq) (\vx - \vy))}
\phi_1(\vx,t)\,\phi_2(\vy,t)\, ,
\en
where
\begin{equation}
\begin{tabular}{|c|c|}
\hline
$i$ & $S_i{\vq}$ \\
\hline 
1 & $(q_1,q_2,q_3)$ \\
2 & $(q_2,-q_1,q_3)$ \\
3 & $(-q_2,q_1,q_3)$ \\
4 & $(-q_1,-q_2,q_3)$ \\
5 & $(-q_2,-q_1,q_3)$ \\
6 & $(q_2,q_1,q_3)$ \\
7 & $(-q_1,q_2,q_3)$ \\
8 & $(q_1,-q_2,q_3)$ \\
\hline
\end{tabular}
\end{equation}
From this expression one readily obtains, for example, the operator
that transforms according to the representation $E$,
\eq
O^E(\vp,\vq,t)=\sum_{\vx, \vy}
e^{i \vp \vx}\,
\left(e^{i \vq_\perp(\vx - \vy)_\perp}-e^{-i \vq_\perp(\vx - \vy)_\perp}\right)\,
e^{i \vq_\parallel(\vx - \vy)_\parallel} \,
\phi_1(\vx,t)\,\phi_2(\vy,t)\, ,
\en
where $\vq_\parallel=(0,0,q_3)$ and $\vq_\perp=(q_1,q_2,0)$, and
similarly for $\vx$, $\vy$.
%In the case of equal masses, the projected operator is given by
%\begin{equation}
%\begin{split}
%O^{E^\pm}(\vp,\vq,t)=\sum_{\vx, \vy}
%&\left(e^{i (\vp \vx+\vq_\parallel(\vx - \vy)_\parallel)} \mp e^{-i (\vp
%  \vx+\vq_\parallel(\vx - \vy)_\parallel)}\right) \\
%&\times \left(e^{i \vq_\perp(\vx - \vy)_\perp}-e^{-i \vq_\perp(\vx -
%\vy)_\perp}\right)\, 
%\phi_1(\vx,t)\,\phi_2(\vy,t)\, ,
%\end{split}
%\end{equation}

\subsubsection{The case $\vd=(1,1,0)$ -- 
product of pion and nucleon fields}

This case is trivial, as only the irreducible representation $G_1$ contributes.
Any operator, for example
\eq
O(\vp,\vq,t)=\sum_{\vx, \vy}
e^{i(\vp \vx + \vq (\vx - \vy))}\,
\pi(\vx,t)\,N(\vy,t)\, ,
\en
will transform according to $G_1$.

Having the characters $\chi(\Gamma)$ of the irreducible representations
$\Gamma$ of the little groups at hand, it should be no problem to
construct operators that transform according to any other
representation. Examples of meson-baryon operators projected onto 
representations $G_2$ in the case of $\vd=(0,0,1)$ and $B_1$, $B_2$ in
the case of $\vd=(1,1,1)$ will be given in a separate
publication~\cite{tbp}, together with numerical results.

\section{Conclusions}
\label{sec7}

In this work we have extended previous work by
L\"uscher~\cite{Luscher} and
others~\cite{Rummukainen:1995vs,Fu:2011xz,Feng:2011ah,Leskovec:2012gb}
on determining the scattering phases from the energy levels of the
(lattice) Hamiltonian in a finite volume to meson and baryon
resonances of arbitrary masses 
and arbitrary total 
momenta $\vP=(2\pi n/L)\,(d_1,d_2,d_3)$ with $d_i=0,\pm 1$, $n \in
\mathbb{Z}$. 
Explicit formulae for the phase shifts have been
given for meson resonances with angular momentum $l \leq 2$ and for
baryon resonances with spin $J \leq 3/2$ and orbital angular momentum
$l \leq 2$. That covers essentially all elastic two-body resonances.
There are several advantages to performing simulations with
nonvanishing total momenta. This includes making the avoided level
crossing in $P$-wave decays occur at a smaller volume, in the case
the scattering particles have different mass, and making a wider set
of energy levels available on a single lattice volume.

The drawback is that the individual partial waves will mix in
general. Neglecting $D$ waves, this is the case for all $S$-wave
meson resonances and all $S$- and $P$-wave spin-$1/2$ baryon
resonances. To compute the $P$-wave phase shift 
$\delta_{\frac{1}{2} 1}$, for example, one will need input from
$\delta_{\frac{1}{2} 0}$. One might be lucky though and find the
latter to be small, because no low-lying positive parity $S$-wave
spin-$1/2$ pion-nucleon resonance has been reported~\cite{pdg}. This
is one of the mysteries of baryon spectroscopy. 

The success of the method depends on our ability to construct operators
that will transform according to the desired representation of the little
group. We have outlined the general procedure of how to construct
such operators from the character tables, and given a few explicit
examples of single-particle and two-particle operators.

% nonvanishing momenta important for baryon P-waves because of the
% small pion mass 

\section*{Acknowledgment}

We like to thank Sasa Prelovsek for discussions. 
This work has been supported in part by the EU Integrated Infrastructure
Initiative {\it HadronPhysics3} under Grant Agreement no. 283286 and
by the DFG under contract SFB/TR 55 (Hadron Physics from Lattice QCD)
and SFB/TR 16 (Subnuclear Structure of Matter), as well as by COSY FFE
under contract no. 41821485 (COSY 106). AR acknowledges support of the
Shota Rustaveli Science Foundation (Project DI/13/02). 

\clearpage

\appendix

\section{Zeta functions}

A valid representation of the zeta function
for $\delta=1$ is given by~\cite{Luscher2} 
\begin{equation}
Z_{lm}^{\vDelta}(1,q^2) = \sum_{\substack{\vz \in P_{\vDelta}\\|\vz|<\lambda}}\,
\frac{\mathcal{Y}_{lm}(\vz)}{\vz^2 - q^2} + 
(2\pi)^3\,\int_0^\infty dt\, \left[e^{tq^2}\,K_{lm}^{{\vDelta}\,
    \lambda}(t,{\bf 0}) -
  \frac{\gamma\,\delta_{l0}\,\delta_{m0}}{(4\pi)^2t^{3/2}}\right] \,,
\end{equation}
where 
\begin{equation}
%\begin{split}
K_{lm}^{\vDelta\,\lambda}(t,\vre) = K_{lm}^{\vDelta}(t,\vre) - \frac{1}{(2\pi)^3}
\sum_{\substack{\vz \in  P_{\vDelta}\\|\vz|<\lambda}}\,
\mathcal{Y}_{lm}(\vz)\, e^{i\vz \vre - t\vz^2} 
\end{equation}
and $K_{lm}^{\vDelta}(t,\vre)$ is the heat kernel of the Laplace
operator on the $\vDelta$-periodic lattice, 
\begin{equation}
K_{lm}^{\vDelta}(t,\vre) = \frac{1}{(2\pi)^3} \sum_{\vz \in
P_{\vDelta}} \mathcal{Y}_{lm}(\vz)\,e^{i\vz \vre - t\vz^2} \,.
\end{equation}
This leads to
\begin{equation}
\begin{split}
Z_{lm}^{\vDelta}(1,q^2) &= \sum_{\substack{\vz \in
  P_{\vDelta}\\|\vz|<\lambda}}\,
\frac{\mathcal{Y}_{lm}(\vz)}{\vz^2 - q^2}\,e^{-(\vz^2-q^2)} + \int_1^\infty dt \,
\sum_{\substack{\vz \in
    P_{\vDelta}\\|\vz|>\lambda}}\,\mathcal{Y}_{lm}(\vz)\,
  e^{-t(\vz^2-q^2)} \\
&+\int_0^1 dt \,\left[ \sum_{\vz \in
    P_{\vDelta}}\,\mathcal{Y}_{lm}(\vz)\, e^{-t(\vz^2-q^2)} -\gamma\,
    \frac{\pi}{2} \, \delta_{l0}\,\delta_{m0}\,\frac{1}{t^{3/2}}\right] -
    \gamma \,\frac{\pi}{4}\, \delta_{l0}\,\delta_{m0} 
\end{split}
\end{equation}
with both integrals being well defined for a suitable choice of
$\lambda$. Indeed, using the relation
\begin{equation}
\sum_{\vz \in
  P_{\vDelta}} \mathcal{Y}_{lm}(\vz)\,e^{- t\vz^2} =
\gamma\left(\frac{\pi}{t}\right)^{3/2}\,\left(\frac{i}{2t}\right)^l \sum_{\vn 
  \in \mathbb{Z}^3} e^{-i\pi \vn\vDelta}
\mathcal{Y}_{lm}(2\pi\vgamma\vn)\,e^{-(2\pi\vgamma\vn)^2/4t} \,,
\end{equation}
the sum over $\vz \in P_{\vDelta}$ in the second integral can be
expressed in terms of a sum over $\vn \in \mathbb{Z}^3$, which
finally gives
\begin{equation}
\begin{split}
Z_{lm}^{\vDelta}(1,q^2) &= \sum_{\vz \in P_{\vDelta}}\,
\frac{\mathcal{Y}_{lm}(\vz)}{\vz^2 - q^2}\,e^{-(\vz^2-q^2)} + \gamma
\,\frac{\pi}{2}\,\delta_{l0}\,\delta_{m0}\, F(q) \\
&+ \gamma \pi^{3/2}\,
\int_0^1 dt \, \frac{e^{t q^2}}{t^{3/2}}\, \left(\frac{i}{2t}\right)^l
\sum_{\substack{\vn \in \mathbb{Z}^3\\\vn \neq 0}} \,e^{-i\pi \vn\vDelta}\,
\mathcal{Y}_{lm}(2\pi \vgamma \vn)\, e^{-(\pi \vgamma \vn)^2/t} \,,
\end{split}
\end{equation}
where 
\begin{equation}
%\begin{split}
F(q)= \int_0^1\, dt \frac{e^{t q^2}-1}{t^{3/2}} - 2
= \sum_{n=0}^\infty \frac{q^{2n}}{(n-1/2)\, n!} \,.
%\end{split}
\end{equation}

%\clearpage


\begin{thebibliography}{99}

\bibitem{Luscher}
  M.~L\"uscher,
  Commun.\ Math.\ Phys.\  {\bf 105} (1986) 153;
  Nucl.\ Phys.\  B {\bf 364}, 237 (1991).

\bibitem{Rummukainen:1995vs} 
  K.~Rummukainen and S.~A.~Gottlieb,
  %``Resonance scattering phase shifts on a nonrest frame lattice,''
  Nucl.\ Phys.\ B {\bf 450}, 397 (1995)
  [hep-lat/9503028].

\bibitem{Bietenholz:2011qq} 
  W.~Bietenholz, V. Bornyakov, M. G\"ockeler, R. Horsley,
  W.G. Lockhart, Y. Nakamura, H.~Perlt, D. Pleiter, P.E.L. Rakow,
  G. Schierholz, A. Schiller, T. Streuer, H. St\"uben, F. Winter,
  J.M. Zanotti [QCDSF Collaboration], 
  %``Flavour blindness and patterns of flavour symmetry breaking in
  %lattice simulations of up, down and strange quarks,'' 
  Phys.\ Rev.\ D {\bf 84}, 054509 (2011)
  [arXiv:1102.5300 [hep-lat]].

\bibitem{Durr:2008zz} 
  S.~D\"urr, Z. Fodor, J. Frison, C. Hoelbling, R. Hoffmann,
  S.D. Katz, S. Krieg, T. Kurth, L.~Lellouch, T. Lippert, K.K. Szabo,
  G. Vulvert [BMW Collaboration],  
  %``Ab-Initio Determination of Light Hadron Masses,''
  Science {\bf 322}, 1224 (2008)
  [arXiv:0906.3599 [hep-lat]].

\bibitem{Davoudi:2011md} 
  Z.~Davoudi and M.~J.~Savage,
  %``Improving the Volume Dependence of Two-Body Binding Energies
  %Calculated with Lattice QCD,'' 
  Phys.\ Rev.\ D {\bf 84}, 114502 (2011)
  [arXiv:1108.5371 [hep-lat]].

\bibitem{Fu:2011xz} 
  Z.~Fu,
  %``Rummukainen-Gottlieb's formula on two-particle system with
  %different mass,'' 
  Phys.\ Rev.\ D {\bf 85}, 014506 (2012)
  [arXiv:1110.0319 [hep-lat]].

\bibitem{Feng:2011ah} 
  X.~Feng, K. Jansen, D.B. Renner [ETM Collaboration],
  %``A new moving frame to extract scattering phases in lattice QCD,''
  PoS LAT {\bf 2010}, 104 (2010)
  [arXiv:1104.0058 [hep-lat]].

\bibitem{Leskovec:2012gb} 
  L.~Leskovec and S.~Prelovsek,
  %``Scattering phase shifts for two particles of different mass and
  %non-zero total momentum in lattice QCD,'' 
  arXiv:1202.2145 [hep-lat].

\bibitem{Doring:2012eu} 
  M.~D\"oring, U.-G. Mei{\ss}ner, E. Oset, A. Rusetsky,
  %``Scalar mesons moving in a finite volume and the role of partial
  %wave mixing,'' 
  arXiv:1205.4838 [hep-lat].

\bibitem{Luscher2}
  M.~L\"uscher,
  Nucl.\ Phys.\  B {\bf 354}, 531 (1991).

\bibitem{Messiah}
  A.~ Messiah, ``Quantum Mechanics, Volume II'', Dover Publications (2000).

\bibitem{Bernard:2008ax}
  V.~Bernard, M. Lage, U.-G. Mei{\ss}ner, A. Rusetsky,
  %``Resonance properties from the finite-volume energy spectrum,''
  JHEP {\bf 0808} (2008) 024
  [arXiv:0806.4495 [hep-lat]].

\bibitem{pdg}
  K. Nakamura {\it et al.} [Particle Data Group], J. Phys. G {\bf 37},
  075021 (2010).

\bibitem{Thomas:2011rh} 
  C.~E.~Thomas, R.~G.~Edwards and J.~J.~Dudek,
  %``Helicity operators for mesons in flight on the lattice,''
  Phys.\ Rev.\ D {\bf 85}, 014507 (2012)
  [arXiv:1107.1930 [hep-lat]].

\bibitem{Dudek:2012gj} 
  J.~J.~Dudek, R.~G.~Edwards and C.~E.~Thomas,
  %``S and D-wave phase shifts in isospin-2 pi pi scattering from lattice QCD,''
  arXiv:1203.6041 [hep-ph].

\bibitem{Foley:2012wb} 
  J.~Foley, J.~Bulava, Y.~-C.~Jhang, K.~J.~Juge, D.~Lenkner,
  C.~Morningstar and C.~H.~Wong, 
  %``Group-theoretical construction of finite-momentum and
  %multi-particle operators for lattice hadron spectroscopy,'' 
  arXiv:1205.4223 [hep-lat].

\bibitem{Elliott}
  J.~P.~Elliott and P.~G.~Dawber,
  ``Symmetry in Physics, Volume I: Principles And Simple Applications'',
  Macmillan (1979).
 
\bibitem{tbp} M.~G\"ockeler, R.~Horsley, M.~Lage, U.-G.~Mei{\ss}ner, R.~Millo,
  A.~Nogga, P.E.L.~Rakow, A.~Rusetsky, G.~Schierholz, J.~Zanotti, work
  in progress.

\end{thebibliography}
\end{document}